\documentclass[ aps,
                physrev,
                preprint,
                onecolumn,
                floatfix,
                superscriptaddress]{revtex4-2}
\usepackage{mlmodern}
\usepackage{graphicx}
\usepackage{bm}
\usepackage{physics}
\usepackage[unicode=true,
    psdextra,
    colorlinks=true,
    pdfstartview=FitV,
    linkcolor=blue,
    citecolor=blue,
    urlcolor=blue,
    anchorcolor = blue]{hyperref}

\begin{document}

% Use the \preprint command to place your local institutional report
% number in the upper righthand corner of the title page in preprint mode.
% Multiple \preprint commands are allowed.
% Use the 'preprintnumbers' class option to override journal defaults
% to display numbers if necessary
%\preprint{}

%%%%%%%%%%%%%%%%%%%%%%%%%%%%%%%% TITLE %%%%%%%%%%%%%%%%%%%%%%%%%%%%%%%%%%%%%%%%%%%%
\title{On Large Deformations of Oldroyd-B Drops in a Steady Electric Field}

% repeat the \author .. \affiliation  etc. as needed
% \email, \thanks, \homepage, \altaffiliation all apply to the current
% author. Explanatory text should go in the []'s, actual e-mail
% address or url should go in the {}'s for \email and \homepage.
% Please use the appropriate macro foreach each type of information

% \affiliation command applies to all authors since the last
% \affiliation command. The \affiliation command should follow the
% other information
% \affiliation can be followed by \email, \homepage, \thanks as well.

%%%%%%%%%%%%%%%%%%%%%%%%%%%%%%%%% AUTHORS %%%%%%%%%%%%%%%%%%%%%%%%%%%%%%%%%%%%%%%%%
\author{Sarika Shivaji Bangar}
\affiliation{Department of Mechanical Engineering, Indian Institute of Science, Bengaluru-560012, India}

\author{Gaurav Tomar}%
\email[Corresponding author: ]{gtom@iisc.ac.in}
\affiliation{Department of Mechanical Engineering, Indian Institute of Science, Bengaluru-560012, India}%

%Collaboration name if desired (requires use of superscriptaddress
%option in \documentclass). \noaffiliation is required (may also be
%used with the \author command).
%\collaboration can be followed by \email, \homepage, \thanks as well.
%\collaboration{}
%\noaffiliation
\date{\today}

%%%%%%%%%%%%%%%%%%%%%%%%%%%%%%%% ABSTRACT %%%%%%%%%%%%%%%%%%%%%%%%%%%%%%%%%%%%%%%%%
\begin{abstract}
The deformation of viscoelastic drops under electric fields is central to applications in microfluidics, inkjet printing, and electrohydrodynamic manipulation of complex fluids. This study investigates the dynamics of an Oldroyd-B drop subjected to a uniform electric field using numerical simulations performed with the open-source solver Basilisk. Representative pairs of conductivity ratio ($\sigma_r$) and permittivity ratio ($\epsilon_r$) are selected from six regions ($PR_A^+$, $PR_B^+$, $PR_A^-$, $PR_B^-$, $OB^+$, and $OB^-$) of the $(\sigma_r, \epsilon_r)$ phase space. In regions where the first- and second-order deformation coefficients share the same sign ($PR_A^-$, $PR_B^-$, $OB^+$), deviations from Newtonian behavior are negligible. In $PR_A^+$, drops develop multi-lobed shapes above a critical electric capillary number, with elasticity reducing deformation and increasing the critical $Ca_E$ with Deborah number ($De$). In $PR_B^+$, drops form shapes with conical ends above the critical $Ca_E$, while steady-state deformation decreases with $De$ below this threshold, and critical $Ca_E$ shows non-monotonic variation. At high $Ca_E$ and $De$, transient deformation exhibits maxima and minima before reaching steady state, with occasional oscillations between spheroidal and pointed shapes. 
In $OB^-$, drops deform to oblate shapes and breakup above a critical $Ca_E$, with deformation magnitude increasing and critical $Ca_E$ decreasing with $De$; at low $Ca_E$ and high $De$, dimpling and positional oscillations are observed. These results elucidate viscoelastic-electric interactions and provide guidance for controlling drop behavior in practical applications.
\end{abstract}

%%%%%%%%%%%%%%%%%%%%%%%%%%%%%%%% KEYWORDS %%%%%%%%%%%%%%%%%%%%%%%%%%%%%%%%%%%%%%%
% insert suggested keywords - APS authors don't need to do this
%\keywords{}
%\maketitle must follow title, authors, abstract, and keywords
\maketitle
\section{Introduction}\label{Sec_Intro}
Understanding the dynamics and deformation of liquid droplets suspended in an ambient fluid under the influence of an electric field is essential for numerous industrial applications, including inkjet printing \cite{basaran2013nonstandard, lau2017ink}, de-emulsification of oil \cite{eow2002electrostatic, alvarado2010enhanced, zhang2011application}, electrospraying and atomization \cite{kelly1984electrostatic, law2018electrostatic}, electrostatic spray painting \cite{hines1966electrostatic}, electric propulsion \cite{moreau2015electrohydrodynamic, huh2019numerical}, and the manipulation of droplets in microfluidic systems \cite{laser2004review, stone2004engineering, phan2025demand}. Moreover, investigating these phenomena also offers valuable insights into natural processes such as rain electrification, droplet breakup during thunderstorms, and atmospheric charge generation \cite{simpson1909electricity, wilson1921iii, blanchard1963electrification}.

When a fluid drop is immersed in a dielectric medium and subjected to an electric field, it typically deforms into a prolate spheroid at low field strengths and undergoes breakup at high fields, as established by \citet{o1953distortion} and \citet{taylor1964disintegration}. \citet{allan1962particle} explained the prolate deformation of conducting drops by balancing the normal electric stresses at the interface with the interfacial tension. \citet{o1953distortion} independently derived this result from energy considerations.
However, \citet{allan1962particle} also observed that even weakly conducting drops deformed to the oblate shapes. To address this, \citet{o1957electric} proposed fluid conductivity as a contributing factor, developing a theoretical model for the drop deformation (characterized by eccentricity) based on energy considerations, though it neglected energy from interfacial charge separation and induced flow.
\citet{taylor1966studies} elucidated the role of tangential electric stresses, showing that it induces internal and external circulations and governs whether a drop deforms into a prolate or oblate shape, depending on the conductivity and permittivity ratios. His leaky-dielectric model accounts for finite conductivities while assuming rapid charge relaxation, thereby decoupling electric and flow fields. The model established the strong dependence of deformation on material property ratios. \citet{melcher1969electrohydrodynamics} further analyzed interfacial shear stresses within this framework. \citet{torza1971electrohydrodynamic} extended it to alternating fields, finding qualitative agreement albeit with  larger experimental deformations than those predicted by theory. To address this, \citet{ajayi1978note} introduced higher-order corrections, which increased predicted deformations but did not fully resolve the discrepancies. A comprehensive review of leaky dielectric model is given in  \citet{saville1997electrohydrodynamics}.

The leaky dielectric model (LDM) has been extensively employed to investigate the electrohydrodynamic behavior of drops in electric fields. \citet{ha1995effects} analyzed steady-state deformation and stability in the presence of surfactants, demonstrating that surfactants enhance deformation, with a stronger effect on prolate than on oblate shapes. Building on the LDM, \citet{bentenitis2005droplet} developed the extended leaky dielectric model (ELDM) to capture continuous deformation and hysteresis at high field strengths, emphasizing that the conductivity and viscosity ratios primarily govern parallel deformations, while the conductivity and permittivity ratios control perpendicular ones.
\citet{zabarankin2013liquid} formulated an analytical model for predicting moderate deformations by approximating the drop as a spheroid. Similarly, \citet{nganguia2013equilibrium} examined the deformation of surfactant-laden drops using a second-order small deformation theory for weak fields and a spheroidal model for larger deformations. \citet{deshmukh2013deformation} explored drops in quadrupole electric fields, revealing that the deformation is dominated by the fourth-order Legendre mode and that the resulting internal flow exhibits complex circulatory structures and distinct deformation–circulation phase characteristics.
The electrorotation behavior of drops has also been studied within the LDM framework. \citet{he2013electrorotation} investigated drop rotation under DC fields using a three-dimensional model, identifying surface charge convection as the key mechanism and linking rotational thresholds to the curious Quincke rotation of rigid spheres. Extending this analysis, \citet{yariv2016electrohydrodynamic} examined drop electrorotation in the asymptotic limit of large electric Reynolds number using a two-dimensional formulation.
The classical Taylor–Melcher framework has further been used to describe symmetry-breaking electrohydrodynamic instabilities such as Quincke rotation, equatorial streaming, and pattern formation on surfactant-laden drops\cite{vlahovska2016electrohydrodynamic, vlahovska2019electrohydrodynamics}. Moreover, surface charge convection has been shown to significantly influence drop breakup modes\cite{sengupta2017role}.

For transient drop deformation, \citet{sozou1972electrohydrodynamics} made the first attempt by incorporating the local acceleration term ($\rho\pdv{\bm{u}}{t}$) into the governing equations. Using a quasi-steady approximation, \citet{moriya1986deformation} analyzed time-dependent deformation in weak electric fields, assuming an extensional flow during the transition from a spherical to a spheroidal shape. \citet{esmaeeli2011transient} later derived a closed-form solution for transient deformation of a leaky-dielectric drop in small fields, quantifying the roles of normal and tangential electric stresses in driving the flow and deformation.
Within the Taylor–Melcher framework, \citet{zhang2013transient} developed a model for transient deformation accounting for finite charge relaxation time while assuming the drop remained spheroidal throughout. \citet{lanauze2013influence} further included both finite charge relaxation and inertia, showing that deformation overshoot is governed by the Ohnesorge number—representing the ratio of capillary to momentum diffusion timescales—while the presence of an initial prolate deformation for oblate final states is determined by the Saville number, the ratio of electric to viscous diffusion timescales.
Using a boundary integral approach, \citet{lanauze2015nonlinear} examined nonlinear transient electrohydrodynamics with surface charge convection and finite charge relaxation, focusing on conditions leading to stable oblate shapes. They demonstrated that slower charge relaxation induces a transient prolate-to-oblate transition due to delayed interfacial charge buildup, and that charge convection toward the equator weakens oblate deformation. \citet{vlahovska2019electrohydrodynamics} later generalized Taylor’s leaky-dielectric model to capture transient drop dynamics.
Most recently, \citet{esmaeeli2020transient} performed three-dimensional simulations at finite Reynolds numbers, revealing that drop deformation evolves at timescale, $\tau_{def}$, with peak deformation governed by the capillary number ($Ca = \tau_{def}/\tau_{conv}$, where $\tau_{def}$ is the viscous capillary time scale and $\tau_{conv}$ is the convective time scale). The long-term behavior depends on the Ohnesorge number ($Oh^2 = \tau_{def}/\tau_{diff}$; where $\tau_{diff}$ is the viscous diffusion time scale) 
%{\bf where $\tau_{diff}$ is ??}) 
for $Oh^2 > 1$, the dynamics resemble a monotonic mass–damper system, while for $Oh^2 < 1$, they exhibit oscillatory, mass–spring–damper behavior. Although steady deformation is independent of $Oh^2$, increasing the Reynolds number (since $Oh^2 = Ca/Re_f$; $Re_f = \tau_{diff}/\tau_{conv}$) enhances oscillatory transients. 
% { \bf define reynolds number here }

Experimental investigations of drop deformation and breakup in electric fields have also been conducted. \citet{nishiwaki1988deformation} studied drop deformation experimentally, providing an empirical relation for retardation time based on fluid viscosities, interfacial surface energy, and drop radius. They found that the theory could predict equilibrium deformations for poly-propylene oxide and polypropylene glycol drops, but water drop deformations were smaller than predicted. \citet{vizika1992electrohydrodynamic} experimentally determined drop deformation under steady and oscillatory electric fields, showing better agreement with leaky dielectric theory compared to  \citet{torza1971electrohydrodynamic}. More recently, \citet{karp2024electrohydrodynamic} investigated internal drop circulation and interfacial dynamics using particle image velocimetry and high-speed shadowgraphy, finding good agreement with the leaky dielectric model for steady and transient deformations of magnitudes less than 0.1.
Experimental investigations of \citet{brosseau2017streaming} demonstrated that low-viscosity drops (viscosity ratio $<$ 0.1) subjected to strong DC electric fields undergo equatorial streaming. The electric shear stresses along the interface drive a surface flow that converges at the equator, where amplification of interfacial perturbations results in the ejection of a fluid sheet from the stagnation line.

Numerical studies have greatly advanced the understanding of Newtonian drop electrohydrodynamics, elucidating deformation, stability, and breakup mechanisms. \citet{miksis1981shape} first employed the boundary integral method to compute static dielectric drop shapes, revealing obtuse-angled conical ends at high field strengths for permittivity ratios above a critical value, and prolate spheroids otherwise. Extending this, \citet{sherwood1988breakup} analyzed leaky dielectric drops under creeping flow, predicting experimentally observed breakup modes such as tip streaming and blob–thread formations. His results linked high permittivity ratios to pointed-end drops and high conductivities to blob division, though the analysis was limited to prolate deformations.
Using a finite element Galerkin approach, \citet{feng1996computational} solved the full electrohydrodynamic problem and showed that asymptotic theories underestimate deformation and flow. They identified a critical electric field beyond which no stable shape exists and found deformation behavior to be strongly dependent on conductivity and viscosity ratios. Prolate drops were dominated by electrostatic stresses, while oblate shapes required hydrodynamic stresses for stability. Their nonlinear model reconciled discrepancies with experiments. \citet{feng1999electrohydrodynamic} later included charge convection effects, showing it weakened hydrodynamic flow, thereby reducing oblate and enhancing prolate deformations.
Building on Sherwood’s framework, \citet{lac2007axisymmetric} explored drop stability and breakup under Stokes flow, covering a wide range of conductivity–permittivity ratios and viscosity contrasts, including for oblate modes. 
\citet{dubash2007behaviour} analyzed inviscid conducting drops in viscous insulating media, capturing steady deformation at low fields and breakup at critical strengths.
\citet{karyappa2014breakup} studied the effect of viscosity on the deformation and breakup of a perfectly conducting drop in a dielectric medium, identifying three axisymmetric shapes prior to breakup (ASPB) modes (lobes, pointed, non-pointed ends) that transform into non-axisymmetric shapes at breakup (NSAB) modes (charged disintegration, regular jet, open jet). 
More recently, \citet{wagoner2021electrohydrodynamics} simulated equatorial streaming of weakly conducting drops in highly conducting, viscous media, demonstrating that the instability arises only when both surface charge convection and diffusion are considered. Using a lattice Boltzmann framework, \citet{wang2024lattice} further captured fingering equatorial streaming region under strong DC fields and significant contrast in electrical conductivity.

Various computational approaches have been proposed by researchers for the solution of multiphase EHD flows. \citet{zhang20052d} presented a multicomponent Lattice Boltzmann method (LBM) for 2D electrohydrodynamics of a drop. \citet{tomar2007two} used a coupled level set and volume of fluid method for dielectric-dielectric and conducting-conducting fluids, proposing a weighted harmonic mean interpolation scheme for property smoothing. %fields and breakup for critical fields. 
\citet{hua2008numerical} proposed a front tracking/volume of fluid method coupled with various electric field models (perfect dielectric, leaky dielectric, constant surface charge) to predict deformation and motion of viscous drops. \citet{paknemat2012numerical} conducted a numerical study of perfect dielectric, perfect conductive, and leaky dielectric systems using the level-set method with a finite difference-based ghost fluid method, simulating tip-streaming and various breakup modes. 
\citet{nganguia2015immersed} developed an Immersed Interface Method to solve for drop dynamics in an electric field.
\citet{das2017electrohydrodynamics} developed a three dimensional boundary element method for complete leaky dielectric model. They have also considered the effect of surface charge convection and studied the regime of symmetric breaking which leads to Quinke rotation.
\citet{liu2019phase} developed a numerical method for electrohydrodynamic flows using LBM to solve the coupled  Allen-Cahn, Poisson, and Navier-Stokes equations.
\citet{wang2024lattice} developed a lattice Boltzmann framework by coupling an Allen–Cahn-type multiphase model with two additional LB equations to solve the Poisson equation for the electric field and the surface charge conservation equation.

Many fluids encountered in natural and industrial processes are non-Newtonian, exhibiting complex rheological behavior due to the presence of macromolecules or suspended microstructures. Examples include biological fluids, polymer solutions, and complex mixtures used in polymer processing, microfluidics, and inkjet printing. These fluids display coupled viscous and elastic responses, resulting in non-linear phenomena such as normal stress differences, stress relaxation, and strain hardening, which significantly affect the interfacial dynamics compared to Newtonian fluids. Studies of viscoelastic drop deformation and breakup in shear and extensional flows have shown that elasticity can delay breakup, enhance shape recovery, and modify the critical capillary number for deformation.
\citet{ramaswamy1999deformation} examined the deformation of a FENE-CR drop in extensional flow, demonstrating that drop shape reflects a balance between viscoelastic tensile stresses along the symmetry axis, which draw the interface inward and suppress tip deformation, and viscoelastic modifications to viscous and pressure stresses, which promote elongation. At high extensibility ($L^2=600$), tensile stresses dominate, resulting in reduced deformation and lower tip curvature relative to the Newtonian case. At lower extensibility ($L^2=144$), viscoelastic effects enhance deformation and increase the tip curvature.
\citet{hooper2001transient} showed that in a steady extensional flow, viscoelastic drops elongate less than Newtonian drops, but stored elastic stresses induce rapid retraction during the relaxation phase. Similarly, \citet{aggarwal2007deformation} studied Oldroyd-B drops in shear flow, finding that viscoelasticity reduces deformation at low capillary numbers, while a non-monotonic trend emerges at higher capillary numbers. Transiently, the drops exhibit a deformation overshoot, and the critical capillary number for breakup increases with Deborah number.
Despite the widespread occurrence of viscoelastic fluids, their electrohydrodynamics has been relatively less explored. \citet{ha1999deformation} examined the influence of viscoelasticity on drop deformation and stability in an electric field by modeling the fluid as a second-order fluid. Extending this work, \citet{ha2000deformation} analyzed the deformation and breakup of Newtonian and non-Newtonian conducting drops, showing that both elastic drops with shear-independent viscosity and viscoelastic drops with shear-rate-dependent viscosity exhibit reduced deformation and enhanced stability. They further observed that elasticity in the continuous phase stabilizes the drop at low viscosity ratios but promotes instability at higher ratios. More recently, \citet{lima2014numerical} numerically investigated the electrohydrodynamic deformation of a Giesekus drop, demonstrating that increasing polymer relaxation time suppresses deformation due to enhanced elastic resistance while simultaneously lowering the drop viscosity due to shear-thinning effects. \citet{zhao2025electrohydrodynamic} investigated the electrohydrodynamic deformation of a viscoelastic drop in shear flow and reported that viscoelasticity reduces both drop deformation and rotational motion, although its influence remains subdominant compared to the electrohydrodynamic effects. \citet{DAS2026105633} derived an approximate expression for the deformation of an Oldroyd-B drop in an electric field under the limit of small Deborah numbers and performed numerical simulations to investigate its behavior at higher Deborah numbers.

A detailed understanding of the electrohydrodynamics of viscoelastic drops remains limited. Extending the work of \citet{lac2007axisymmetric} on Newtonian systems, the present study numerically examines the large deformation and breakup behavior of a viscoelastic drop in a uniform electric field using the Oldroyd-B model. The Oldroyd-B formulation represents a constant-viscosity viscoelastic fluid, capturing elastic effects through relaxation dynamics but assuming infinite polymer extensibility. It provides a baseline for assessing the influence of elasticity on electrohydrodynamic deformation. The formulation of the problem, along with the governing equations and boundary conditions, is presented in \S\ref{Sec_Formualtion}. The numerical methodology and grid independence study are described in \S\ref{Sec_Numerics}. The phase plot in the $(\sigma_r, \epsilon_r)$ plane, along with the selected pairs of conductivity and permittivity ratios used in the present study, are shown in \S\ref{Sec_parameterSpace}, followed by the results and discussion in \S\ref{Sec_results}.

\section{Problem formulation}\label{Sec_Formualtion}
\begin{figure*}
	\centering\includegraphics[width=0.5\textwidth]{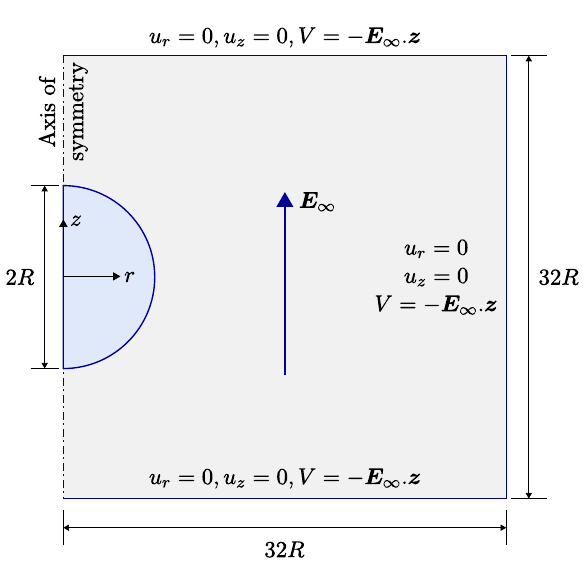}
	\caption{Schematic of the problem setup:  An Oldroyd-B drop of radius $R$ is subjected to an externally applied electric field, $\bm{E}_\infty$, aligned along the axis of symmetry marked in the schematic. The domain size is chosen to be $32R \times 32R$ to minimize the boundary effects.}
	\label{fig_schematic}
\end{figure*}

We consider a viscoelastic drop of radius $R$, suspended in a Newtonian fluid, and subjected to a uniform, steady electric field $\bm{E}_{\infty}$. Subscripts $i$ and $e$ denote the drop and the surrounding medium, respectively. The fluid properties are defined as follows: density $\rho$, viscosity $\mu$, permittivity $\epsilon$, and conductivity $\sigma$. The ambient phase follows the Newtonian constitutive law, while the drop phase is modeled using the Oldroyd-B constitutive relation to account for viscoelastic effects. The viscoelastic response of the drop is characterized by the relaxation time $\lambda_i$, solvent viscosity $\mu_{i_s}$, and polymeric viscosity $\mu_{i_p}$.  
Axisymmetric numerical simulations are performed using the open-source solver Basilisk (\cite{popinet2015quadtree}).
% {\bf Cite Popinet's 2015 paper on basilisk instead of website}.% \href{https://www.basilisk.fr/}{(https://www.basilisk.fr/)}.
A schematic of the configuration is shown in \autoref{fig_schematic}.

\subsection{Governing equations}
The governing equations for the flow field consist of the continuity and momentum balance for an incompressible fluid,
\begin{equation}
\bm{\nabla} \cdot \bm{u} = 0,
\end{equation}
\begin{equation}
\rho \frac{D \bm{u}}{D t} = -\bm{\nabla} p + \bm{\nabla} \cdot \bm{\tau} + \bm{F},
\end{equation}
where $\bm{u}$ is the velocity vector, $p$ is pressure, $\bm{\tau}$ is the stress tensor, and $\bm{F}$ denotes an  external body force. 
  
The Oldroyd-B constitutive relation is used for the drop phase. The total stress is the sum of the solvent and polymeric contributions:
\begin{equation}
\bm{\tau} = \bm{\tau}_s + \bm{\tau}_p,
\end{equation}
with solvent stress $\bm{\tau}_s = 2 \mu_s \bm{D}$, where $\bm{D} = \tfrac{1}{2} \left( \bm{\nabla}\bm{u} + (\bm{\nabla}\bm{u})^T \right)$ is the rate-of-deformation tensor. The polymeric stress obeys the constitutive equation,
\begin{widetext}
\begin{equation*}
\bm{\tau}_p + \lambda \left( \frac{\partial \bm{\tau}_p}{\partial t} + \bm{u} \cdot \bm{\nabla} \bm{\tau}_p - \bm{\tau}_p \cdot \bm{\nabla}\bm{u} - (\bm{\nabla}\bm{u})^T \cdot \bm{\tau}_p \right) = \mu_p \left( \bm{\nabla}\bm{u} + (\bm{\nabla}\bm{u})^T \right),
\end{equation*}
\end{widetext}
where $\lambda$ is the polymer relaxation time, and $\mu_p = \mu - \mu_s$ is the polymeric viscosity.  

In the presence of an electric field $\bm{E}$, the external force is given by the divergence of the Maxwell stress tensor,
\begin{equation}
\bm{F} = \bm{\nabla} \cdot \bm{\tau}_E,
\end{equation}
\begin{equation}
\bm{\tau}_E = \epsilon \left( \bm{E}\bm{E} - \tfrac{1}{2} (\bm{E}\cdot \bm{E}) \bm{I} \right),
\end{equation}
where $\epsilon$ is the permittivity of the medium.
Neglecting magnetic effects, as the characteristic magnetic response time $t_M = \mu_M \sigma L^2$ ($\mu_M$: magnetic permeability, $\sigma$: conductivity, $L$: characteristic length) is typically much smaller than the electric relaxation time $t_E = \epsilon/\sigma$, the electric field is governed by Gauss’s and Faraday’s laws:
\begin{equation}
\bm{\nabla} \cdot (\epsilon \bm{E}) = q, \quad \bm{\nabla} \times \bm{E} = 0,
\end{equation}
where $q$ is the volumetric charge density. Electric field can be represented as a scalar gradient of electric potential $V$, $\bm{E} = -\nabla V$. Thus,
\begin{equation}\label{Eqn_Poisson_potential}
\bm{\nabla}\cdot(\epsilon \bm{\nabla} V) = -q.
\end{equation}
Charge conservation is expressed as
\begin{equation}\label{Eqn_Charge_Conservation}
\frac{\partial q}{\partial t} + \bm{\nabla}\cdot \bm{J} = 0,
\end{equation}
with current density
\begin{equation}
\bm{J} = \sigma \bm{E} + q\bm{u},
\end{equation}
where the first term corresponds to Ohmic conduction and the second term is due to convection of free charges. Equations \eqref{Eqn_Poisson_potential}–\eqref{Eqn_Charge_Conservation} are solved to obtain the electric potential, from which the Maxwell stress and electric body force can be readily computed.  

\subsection{Boundary conditions}
The computational domain is defined in $(r,z)$ coordinates, centered on the drop. The extent of the domain is $-L/2 \leq z \leq L/2$ and $0 \leq r \leq L$ (where $L = 32R$ in the present simulations). At the far-field boundaries, no-slip conditions are imposed,
\begin{subequations}
\begin{equation}
    u_r(r, -L/2) = u_z(r, -L/2) = 0, 
\end{equation}
\begin{equation}
    u_r(r, L/2) = u_z(r, L/2) = 0, 
\end{equation}
\begin{equation}
    u_r(L, z) = u_z(L, z) = 0.
\end{equation}    
\end{subequations}
The applied electric field corresponds to the potential distribution $V = -E_\infty z$. Thus, the boundary conditions for $V$ are,
\begin{subequations}
\begin{equation}
    V(r, -L/2) = -E_\infty z \; ,
\end{equation}
\begin{equation}
    V(r, L/2) = -E_\infty z \; ,
\end{equation}
\begin{equation}
    V(L, z) = -E_\infty z \; .
\end{equation}
\end{subequations}
Polymeric stress components satisfy homogeneous Neumann boundary conditions at the outer boundaries, which introduces only localized errors and do not affect the global solution \citep{alves2021numerical}. Symmetry conditions are imposed along $r=0$:
\begin{subequations}
\begin{equation}
u_r(0,z) = 0, \quad \pdv{u_z}{r}\Big|_{(0,z)} = 0, \quad \pdv{V}{r}\Big|_{(0,z)} = 0,
\end{equation}
\end{subequations}
and for the polymeric stress,
\begin{equation}
     \pdv{\tau_{p_{rr}}}{r}\Big|_{(0, z)} = \pdv{\tau_{p_{zz}}}{r}\Big|_{(0, z)} = \tau_{p_{rz}}(0, z) = \pdv{\tau_{p_{\theta\theta}}}{r}\Big|_{(0, z)} = 0.
\end{equation}

\subsection{Non-dimensional parameters}
The system is non-dimensionalized using $R$ as the length scale and $E_\infty$ as the electric field scale. The velocity scale follows from the balance of electric and viscous stresses,
\begin{equation}
U \sim \frac{\epsilon_e E_\infty^2 R}{\mu_e}.
\end{equation}
Accordingly, stresses and pressure are scaled by $\mu_e U/R$, equivalent to $\epsilon_e E_\infty^2$. The resulting dimensionless groups governing the problem are:
\begin{enumerate}  %[\label=\roman*]
\renewcommand{\theenumi}{\roman{enumi}}
\renewcommand{\labelenumi}{(\theenumi)}
    \item Permittivity ratio, $\epsilon_r = \frac{\epsilon_i}{\epsilon_e}$
    \item Conductivity ratio, $\sigma_r = \frac{\sigma_i}{\sigma_e}$
    \item Density ratio, $\rho_r = \frac{\rho_i}{\rho_e}$
    \item Viscosity ratio, $\mu_r = \frac{\mu_i}{\mu_e}$
    \item Flow Reynolds number, $Re = \frac{\rho_e U R}{\mu_e}$ 
    \item Electric capillary number, $Ca_E = \frac{\epsilon_e E_\infty^2 R}{\gamma}$, is the ratio of electric force to surface tension force with the surface tension being a resisting force and the force due to the electric field being a deforming force.
    \item Deborah number, $De = \frac{\lambda_i U}{R}$, is the ratio of the polymer relaxation time to the flow time scale.
    \item Ratio of solvent to total viscosity of the drop phase, $\beta_i = \frac{\mu_{i_s}}{\mu_i}$
\end{enumerate}

\section{Numerical simulations}\label{Sec_Numerics}
\subsection{Numerical methodology}
For our axisymmetric numerical simulations, we employ a geometric Volume of Fluid (VOF) method in the open-source solver Basilisk. Basilisk employs a second-order accurate time-splitting projection scheme with variables stored at the cell centers. The diffusion term is handled implicitly, while the advection term is calculated using the second-order accurate Bell-Collela-Glaz (BCG) scheme. 
The Poisson equation for the electric potential and the charge conservation equation are solved following the numerical procedure described by \citet{lopez2011charge}. For the viscoelastic stress evolution, Basilisk employs the log-conformation tensor formulation proposed by \citet{fattal2004constitutive} to ensure numerical stability at high Weissenberg numbers.
The polymeric stress is represented in terms of symmetric positive definite tensor, $\bm{A}$, where, $\bm{\tau_p} = \frac{\mu_p}{\lambda}(\bm{A} - \bm{I})$. Equation for $\bm{\Psi} = \log\bm{A}$ is thus given by, 
\begin{equation}\label{equ_Psi}
	\pdv{\bm{\Psi}}{t} + \underbrace{\bm{u}.\bm{\nabla}\bm{\Psi}}_{\text{Advection}} = \underbrace{\bm{\Omega}\bm{\Psi} - \bm{\Psi}\bm{\Omega} + 2\bm{B}}_{\text{Upper convection}} + \underbrace{\frac{1}{\lambda}(e^{-\bm{\Psi}} - \bm{I})}_{\text{Model term}}.
\end{equation}
where $\bm{\Omega}$ is an anti-symmetric tensor and $\bm{B}$ is a symmetric tensor that commutes with $\bm{A}$.
Equation for $\bm{\Psi}$ is solved numerically by time-split procedure  (\citet{lopez2019adaptive}), in which the evolution equation for $\bm{\Psi}$ is split as:
\begin{align}
	&\pdv{\bm{\Psi}}{t} = \bm{\Omega}\bm{\Psi} - \bm{\Psi}\bm{\Omega} + 2\bm{B} \label{Eqn_Psi_upperConv}\\
	&\pdv{\bm{\Psi}}{t} + \grad.(\bm{u\Psi}) = 0 \label{Eqn_Psi_Advection}\\
    &\pdv{\bm{A}}{t} = \frac{1}{\lambda}(\bm{I} - \bm{A}) \label{Eqn_Psi_Model}
\end{align}
The upper-convected derivative (\autoref{Eqn_Psi_upperConv}) is computed explicitly, advection of $\bm{\Psi}$ (\autoref{Eqn_Psi_Advection}) is handled using the BCG scheme , and the model term for Oldroyd-B fluid (\autoref{Eqn_Psi_Model}) is integrated analytically after converting $\bm{\Psi}$ to $\bm{A}$.

\begin{figure*}[ht!]
    \centering
    \includegraphics{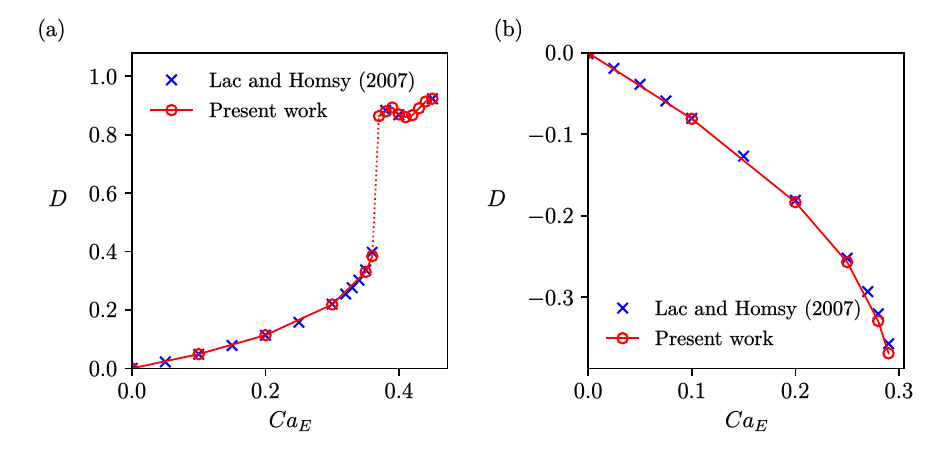}
    \caption{Variation of drop deformation with electric capillary number for $\mu_r=1$, $\rho_r=1$, $Re=1$, $\beta_i = 1$, $De = 0$, (a) $\sigma_r = 10$, $\epsilon_r=1.37$; (b) $\sigma_r = 0.1$, $\epsilon_r=2$}
    \label{Fig_LacAndHomsyValidation}
\end{figure*}
To validate the simulation methodology, we simulate a Newtonian drop ($\beta_i = 1$, $De = 0$) suspended in a Newtonian medium, subjected to a uniform and steady electric field, and compare the deformation results with those of \citet{lac2007axisymmetric}. Simulations are performed keeping the parameters $\mu_r=1$, $\rho_r=1$, $Re=1$, $\beta_i = 1$, $De = 0$ fixed. 
Two pairs of $(\sigma_r, \epsilon_r)$, corresponding to a prolate deformation $(\sigma_r, \epsilon_r)=(10, 1.37)$ and an oblate deformation $(\sigma_r, \epsilon_r)=(0.1, 2)$, are considered. Simulations are performed for various values of $Ca_E$ for each pair of $(\sigma_r, \epsilon_r)$. The drop deformation is quantified using Taylor’s deformation parameter,
\begin{equation}
    D = \frac{L-B}{L+B}
\end{equation}
where $L$ and $B$ denote the drop dimensions along and perpendicular to the applied electric field, respectively. The deformation of the drop, $D$, 
with variation in $Ca_E$ presented in \autoref{Fig_LacAndHomsyValidation} shows good agreement with the results of \citet{lac2007axisymmetric}, thus validating the accuracy and reliability of the numerical approach.

\subsection{Domain independence and grid convergence study}
\begin{figure*}[ht!]
    \centering
    \includegraphics{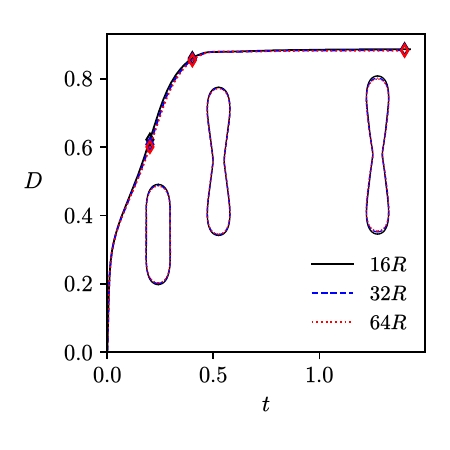}
    \caption{Deformation vs. time for various sizes of the simulation domain. Simulation parameters are $\mu_r=1$, $\rho_r=1$, $Re=1$, $\beta_i = 1/9$, $\sigma_r = 10$, $\epsilon_r=1.37$, $De=5$, $Ca_E=0.4$. $R/\Delta x_{min}$ is taken as 256.}
    \label{Fig_domain_test}
\end{figure*}
A domain independence test is carried out to choose appropriate domain size, as shown in \autoref{Fig_domain_test}. Simulations are performed with three different domain sizes, and the temporal evolution of the drop deformation parameter is compared. The corresponding interface morphologies at three representative time instants are also presented for each domain size. The results demonstrate that the deformation dynamics for domains of size $32R$ and $64R$ are nearly identical, with negligible discrepancies. Based on this observation, a domain size of $32R$ is adopted for all subsequent simulations.

\begin{figure*}[ht!]
	\centering\includegraphics{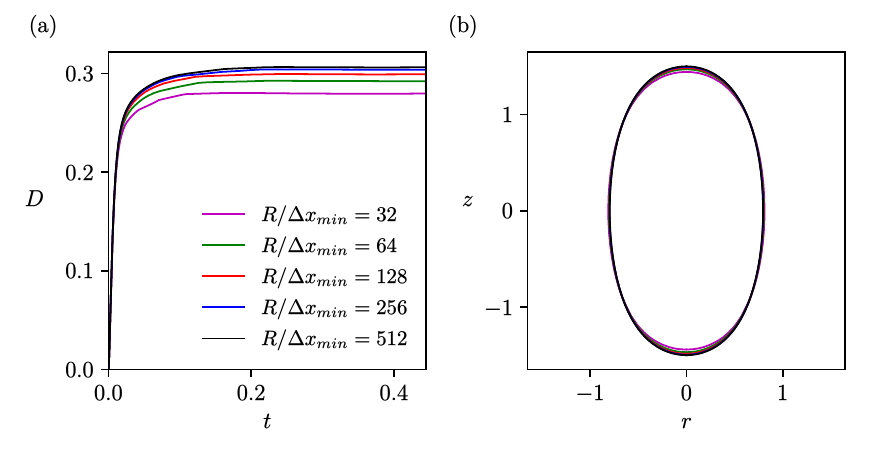}
	\caption{(a) Deformation parameter variation with time at various refinements of the grid. (b) Steady state deformed interface of the drop for various grid refinements. 
    Simulation parameters: $Re=1$, $Ca_E=0.4$, $De=5$, $\sigma_r=10$, $\epsilon_r=1.37$, $\beta=1/9$}
	\label{Fig_gridConvergence}
\end{figure*}

\begin{table}
\centering\caption{The deformation parameter and relative percentage error for different grid refinements}
\label{Table_gridConvergence}
\vspace{1em}
\fontsize{11}{20}\selectfont
\begin{ruledtabular}
\begin{tabular}{ccc}
$R/\Delta x_{min}$ & $D$        & $e_{rel} = \frac{|D - D_{512}|}{|D_{512}|} \times 100$ \\ \hline
32                 & 0.27962977 & 8.674150256       \\
64                 & 0.29198353 & 4.639466719       \\
128                & 0.2991283  & 2.306016345       \\
256                & 0.30390389 & 0.746329711       \\
512                & 0.30618907 & 0                     
\end{tabular}
\end{ruledtabular}
\end{table}

We use adaptive mesh refinement 
% {\bf 
based on the gradient of the void fraction field
% } 
and grid refinement is characterized by the parameter $R/\Delta.x_{min}$. \autoref{Fig_gridConvergence} presents the results of the grid convergence study.
\autoref{Fig_gridConvergence}(a) depicts the temporal evolution of the deformation parameter for different grid resolutions, showing that the results for $R/\Delta x_{min}=512$ and $R/\Delta x_{min}=256$ nearly overlap ($< 1\%$ difference), indicating grid-independence.
\autoref{Fig_gridConvergence}(b) shows the steady-state interface shapes corresponding to various grid refinements, further confirming convergence.
The computed deformation parameters and corresponding relative percentage errors for different grid resolutions are summarized in \autoref{Table_gridConvergence}.
Since the relative percentage error for $R/\Delta x_{min}=256$ is below 1\%, this grid resolution is adopted for all subsequent numerical simulations in the present study.

\section{Parameter space}\label{Sec_parameterSpace}
\begin{figure*}[ht!]
    \centering
    \includegraphics{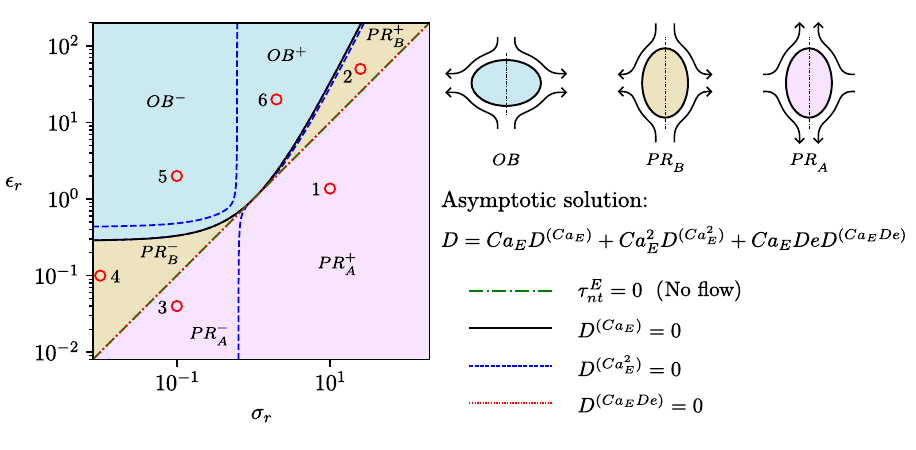}
    \caption{$\epsilon_r-\sigma_r$ phase plot on log-log scale for $\mu_r=1$. $PR$ and $OB$ indicate the prolate and oblate deformation respectively. Subscripts $A$ and $B$ for prolate deformation correspond to the flow direction from equator to poles and poles to equator respectively. Oblate deformation is always accompanied by the poles to equator flow around the drop. Superscript $+$ indicates the positive contribution to the deformation at $\order{Ca_E^2}$, whereas superscript $-$ denotes the negative contribution to the deformation at $\order{Ca_E^2}$. $(\sigma_r, \epsilon_r)$ pairs selected for the study are marked (red circle markers) on $(\sigma_r, \epsilon_r)$ phase plot on log-log scale. Expressions for the deformation coefficients at various orders are given in Appendix \ref{appA}.}
    \label{Fig_CasesSelection}
\end{figure*}
 From asymptotic analysis, valid in the limit of small electric capillary number $(Ca_E)$ and small Deborah number $(De)$, the deformation is found to depend strongly on the conductivity ratio $(\sigma_r)$ and the permittivity ratio $(\epsilon_r)$ \cite{DAS2026105633}. This dependence is represented in the log–log phase plot on the $(\sigma_r, \epsilon_r)$ plane shown in \autoref{Fig_CasesSelection}.% (\citet{lac2007axisymmetric}).
The tangential component of the electric stress at the interface scales with $(\epsilon_r - \sigma_r)$. Consequently, the dash–dotted green line, $\epsilon_r = \sigma_r$, corresponds to vanishing tangential stress. Along this line, the normal electric stress is balanced by surface tension through prolate deformation, while both phases remain quiescent. For $\epsilon_r < \sigma_r$, the induced flow is directed from the equator toward the poles, whereas for $\epsilon_r > \sigma_r$, the flow reverses direction, moving from the poles toward the equator.
The solid black line, defined by $D^{(Ca_E)}=0$, separates the regions of prolate and oblate deformation. Oblate deformation is always  associated with pole-to-equator flow, while prolate deformation may correspond to either equator-to-pole or pole-to-equator flow. The dashed blue line marks the locus where the second-order deformation correction $D^{(Ca_E^2)}$ vanishes.
% {\bf where's the blue line in the figure? or green and red lines??? Legend doesn't seem to match with the lines in the graph.. also the capture needs explanation of the figure... the current line in the caption should be the last line of the caption..}. 
Here, a superscript ‘$+$’ indicates a positive $\mathcal{O}(Ca_E^2)$ correction, and a superscript ‘$-$’ indicates a negative one. Furthermore, since $D^{(Ca_EDe)} \propto (\epsilon_r - \sigma_r)^2$, the red dotted line corresponding to $D^{(Ca_EDe)}=0$ coincides with the no-flow condition $\epsilon_r = \sigma_r$. 
% {\bf 
% in figure 5 don't say electrostatics in the legend corresponding to $\tau_{nt}= 0$.. just say No flow.. 
% the expressions for all these should be put in the appendix.. and then refer to IJMF paper for details..}

In this study, the electrohydrodynamic deformation of a viscoelastic drop is investigated for $\rho_r=1$, $\mu_r=1$, $Re=1$, and $\beta_i=1/9$. Six distinct regions on the $(\sigma_r, \epsilon_r)$ plane, denoted $PR_A^+$, $PR_B^+$, $PR_A^-$, $PR_B^-$, $OB^-$, and $OB^+$, identified from asymptotic analysis, serve as the basis for selection of pairs $(\sigma_r, \epsilon_r)$. A representative point is chosen from each region, as listed below and indicated in \autoref{Fig_CasesSelection}:
\begin{enumerate} %[\label = (\roman*)]
\renewcommand{\theenumi}{\roman{enumi}}
\renewcommand{\labelenumi}{(\theenumi)}
    \item $PR_A^+$: $(\sigma_r, \epsilon_r) = (10, 1.37)$
    \item $PR_B^+$: $(\sigma_r, \epsilon_r) = (25, 50)$
    \item $PR_A^-$: $(\sigma_r, \epsilon_r) = (0.1, 0.04)$
    \item $PR_B^-$: $(\sigma_r, \epsilon_r) = (0.01, 0.1)$
    \item $OB^-$: $(\sigma_r, \epsilon_r) = (0.1, 2)$
    \item $OB^+$: $(\sigma_r, \epsilon_r) = (2, 20)$
\end{enumerate}

For each selected pair, numerical simulations are performed over a range of $Ca_E$ and $De$. The dimensional parameters are fixed as: drop radius $R=0.001$, ambient density $\rho_e = 1000 kg/m^3$, ambient permittivity $\epsilon_e = 4.425 \times 10^{-11}$ farads per $m$, ambient conductivity $\sigma_e = 10^{-5}$ Siemens per $m$, and interfacial tension $\gamma = 0.065 N/m$. Remaining dimensional parameters are obtained from the specified non-dimensional values. The velocity scale, obtained from the balance of electric and viscous stresses, is $U = 0.2549 \sqrt{Ca_E}~~ m/s$ resulting in an electric Reynolds number $Re_E = \frac{\epsilon_e U}{R\sigma_e} = 1.128 \times 10^{-3}\sqrt{Ca_E}$.

\begin{figure*}[ht!]
    \centering
    \includegraphics[width=0.8\linewidth]{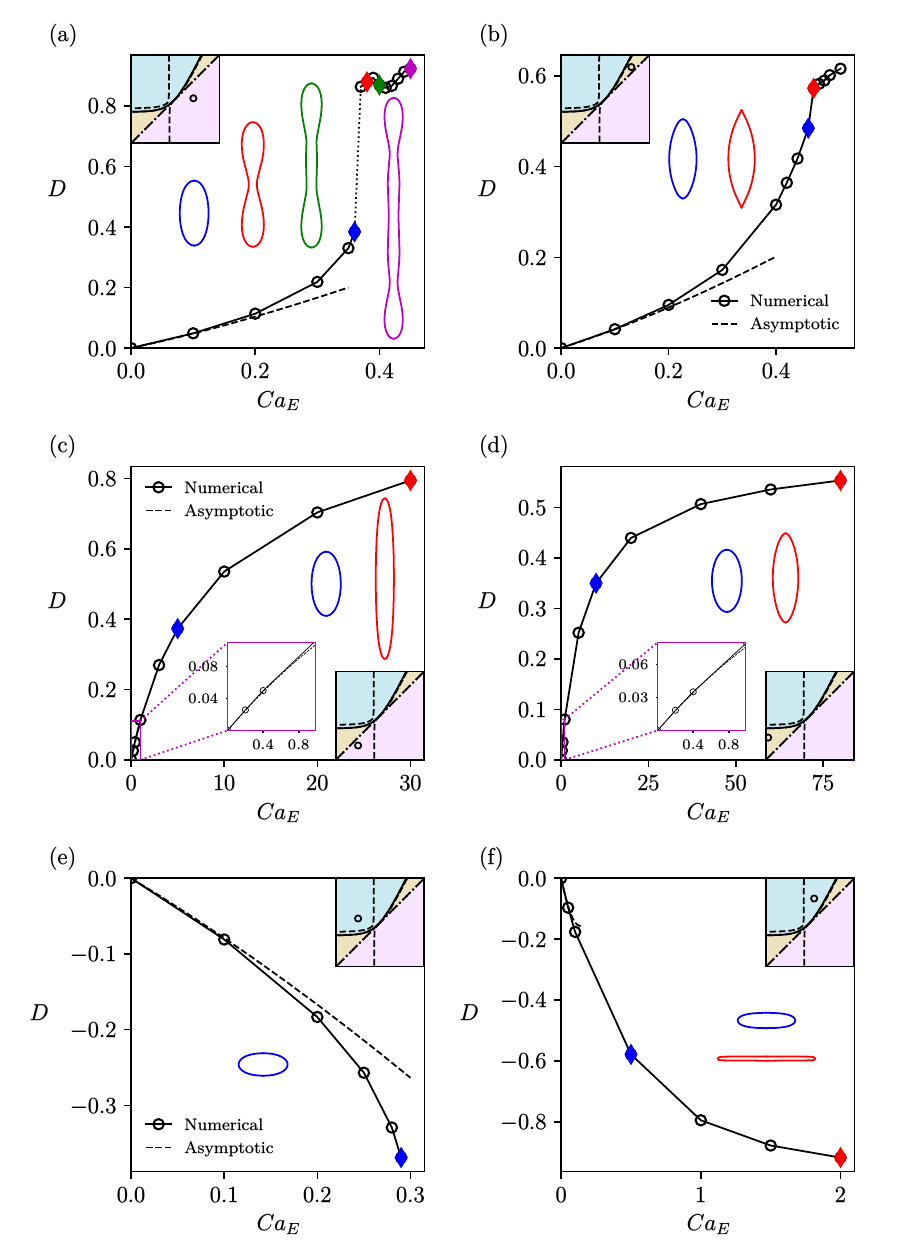}
    \caption{The variation in deformation of a Newtonian drop with electric capillary number $Ca_E$ for $\rho_r=1$, $\mu_r=1$, $Re=1$, (a) $\sigma_r = 10$, $\epsilon_r = 1.37$ (b) $\sigma_r = 25$, $\epsilon_r = 50$ (c) $\sigma_r = 0.1$, $\epsilon_r = 0.04$ (d) $\sigma_r = 0.01$, $\epsilon_r = 0.1$ (e) $\sigma_r = 0.1$, $\epsilon_r = 2$ (f) $\sigma_r = 2$, $\epsilon_r = 20$. Interface shapes correspond to the steady state shape of the drop for the point shown by the diamond marker of same color. Inset shows the location of $(\sigma_r, \epsilon_r)$ point in phase plot. Dashed line shows the drop deformation obtained from asymptotic analysis.}
    \label{Fig_NewtonianBehavior}
\end{figure*}

\section{Results and discussion}\label{Sec_results}
\subsection{Newtonian drops suspended in a Newtonian fluid}
\autoref{Fig_NewtonianBehavior} summarizes the deformation of Newtonian drops across the six representative $(\sigma_r,\epsilon_r)$ pairs under an externally applied electric field. In the $PR_A^+$ and $PR_B^+$ regimes (\autoref{Fig_NewtonianBehavior}(a–b)), the deformation curve exhibits a positive curvature, in agreement with the asymptotic theory. Drops deform into prolate spheroidal shapes up to a critical electric capillary number $Ca_E^{crit}$, beyond which steady spheroids are no longer sustained. The $PR_A^+$ case undergoes a sequence of shape transitions, from spheroidal to multi-lobed configurations, stabilized by internal electrohydrodynamic recirculation as explained by \citet{lac2007axisymmetric}, whereas the $PR_B^+$ case develops pointed ends without tip streaming due to pole-to-equator flow. In contrast, the $PR_A^-$ and $PR_B^-$ regimes (\autoref{Fig_NewtonianBehavior}(c–d)) exhibit negative curvature, such that deformation grows more slowly with $Ca_E$ and eventually plateaus at large $Ca_E$. This trend enables stable drop shapes to persist at high $Ca_E$, with stronger deformation observed in $PR_A^-$ owing to pole-to-equator circulation. For the $OB^-$ regime (\autoref{Fig_NewtonianBehavior}(e)), the oblate deformation grows at an accelerated rate with $Ca_E$, leading to a critical threshold $Ca_E^{crit}$ beyond which breakup occurs due to destabilizing electric pressure at the poles. Finally, in the $OB^+$ regime (\autoref{Fig_NewtonianBehavior}(f)), oblate deformation increases with $Ca_E$ but at a diminishing rate, allowing stable steady shapes even at large $Ca_E$.

\subsection{Viscoelastic drops in a Newtonian fluid}
\begin{figure*}[ht!]
    \centering
    \includegraphics{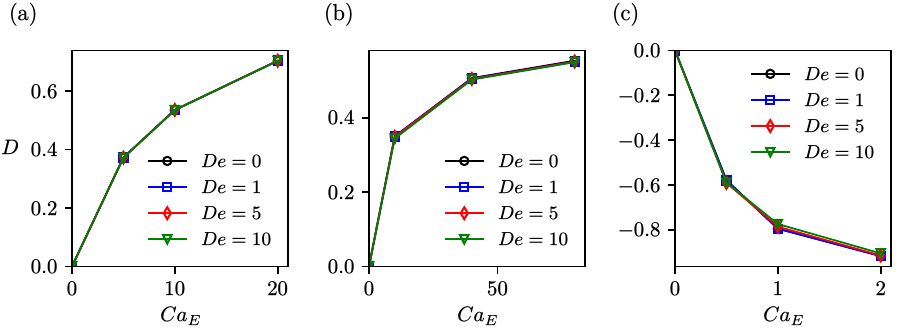}
    \caption{Deformation versus $Ca_E$ for various $De$, for $\rho_r=1$, $\mu_r=1$, $Re=1$, $\beta_i = 1/9$. (a) $\sigma_r=0.1$, $\epsilon_r=0.04$ $(PR_A^-)$ (b) $\sigma_r=0.01$, $\epsilon_r=0.1$ $(PR_B^-)$ (c) $\sigma_r=2$, $\epsilon_r=20$ $(OB^+)$.}
    \label{fig_neg_D1D2}
\end{figure*}
Building on the Newtonian drop analysis of \citet{lac2007axisymmetric} ($De=0$), we extend the investigation to viscoelastic drops subjected to steady electric fields. The classification of $(\sigma_r, \epsilon_r)$ pairs is based on the signs of the first $(D^{(Ca_E)})$ and second $(D^{(Ca_E^2)})$ order deformation coefficients. When both coefficients share the same sign, deformation grows at an accelerated rate with $Ca_E$, giving rise to a critical $Ca_E$ beyond which a steady spheroidal shape cannot be maintained. If the coefficients have opposite signs, deformation increases smoothly with $Ca_E$, and the drop remains stable even at large $Ca_E$.

\begin{figure*}[ht!]
    \centering
    \includegraphics{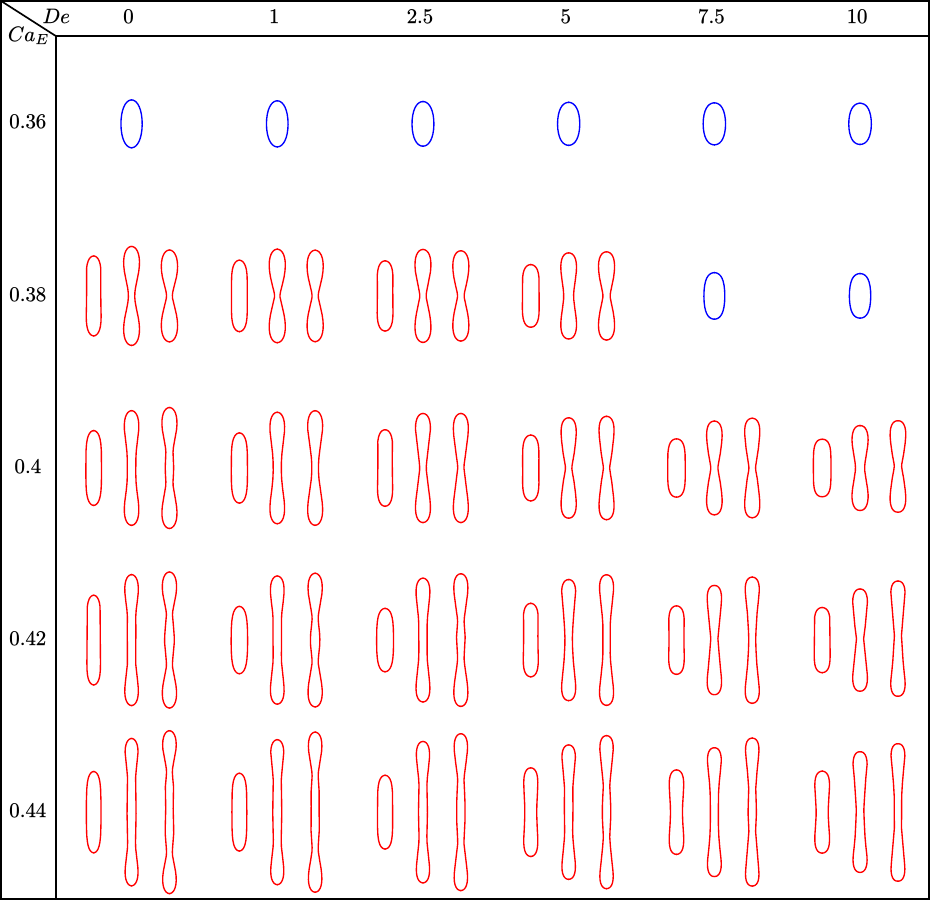}
    \caption{Behavior of an Oldroyd-B drop interface in the presence of an externally applied electric field for various values of $Ca_E$ and $De$ for $(\sigma_r, \epsilon_r) = (10, 1.37)$, $\mu_r=1$, $\rho_r=1$, $Re=1$, $\beta_i=1/9$.}
    \label{Fig_interface_PRA_plus}
\end{figure*}
In the $PR_A^+$ and $PR_B^+$ regions, both coefficients are positive, resulting in prolate deformation with positively curved $D$–$Ca_E$ characteristics. In the $OB^-$ region, both coefficients are negative, yielding oblate deformation with accelerated growth. These three regions ($PR_A^+$, $PR_B^+$, $OB^-$) therefore exhibit a critical $Ca_E$. By contrast, in the $PR_A^-$, $PR_B^-$, and $OB^+$ regions, the coefficients have opposite signs, and deformation evolves smoothly, allowing stable drop shapes even at large $Ca_E$. \autoref{fig_neg_D1D2} shows the variation of drop deformation with electric capillary number for various values of $De$ for $PR_A^-$, $PR_B^-$ and $OB^+$ regions. It is observed that deviation from Newtonian behavior is negligible for these regions. Thus, for subsequent discussions, we focus on $PR_A^+$, $PR_B^+$ and $OB^-$ regions of $(\sigma_r, \epsilon_r)$ plane.

\subsubsection{\texorpdfstring{$(\sigma_r, \epsilon_r)$ from $PR_A^+$ region}{Lg}}
\label{Sec_PRA_plus}

Similar to Newtonian drops, an Oldroyd-B drop deforms into a stable spheroidal shape as long as the electric capillary number ($Ca_E$) remains below a critical threshold. Beyond this threshold, the drop transitions into stable multi-lobed equilibrium configurations.
\autoref{Fig_interface_PRA_plus} demonstrates the evolution of the Oldroyd-B drop interface under a uniform electric field. Across all investigated combinations of $Ca_E$ and $De$, the drop eventually attains a stable equilibrium configuration. For low Deborah numbers ($De = 0, 1, 2.5, 5$), the drop develops a two-lobed shape at $Ca_E = 0.38$, whereas for higher Deborah numbers ($De = 7.5, 10$), this transition occurs at a slightly larger $Ca_E = 0.4$. This shift indicates that the critical electric capillary number increases with elasticity.
Furthermore, the results highlight two complementary trends. For a fixed $De$, both the extent of elongation and the number of lobes in the equilibrium drop shape increase with $Ca_E$, reflecting stronger deformation under higher electric stresses. Conversely, for a fixed $Ca_E$, increasing $De$ reduces the overall elongation and suppresses the formation of multiple lobes, signifying the resistance of viscoelastic stresses to electric-field-induced deformation.

\begin{figure*}[ht!]
    \centering
    \includegraphics{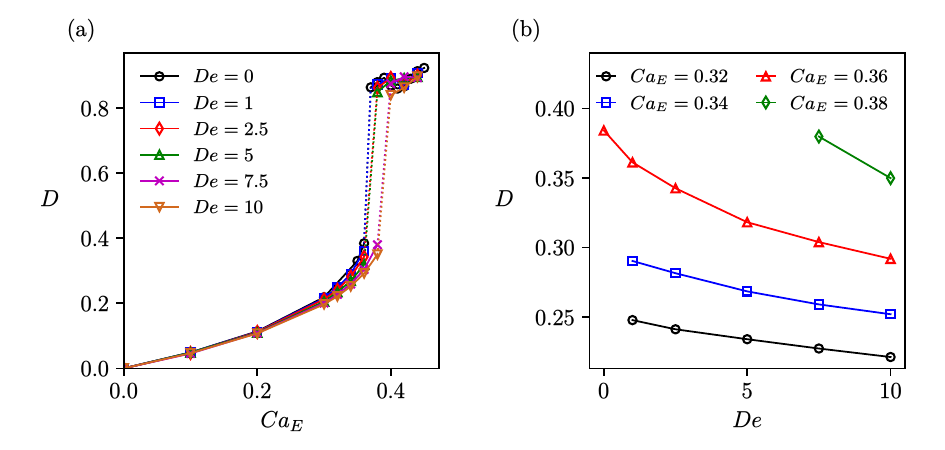}
    \caption{(a) Deformation versus $Ca_E$ for various $De$, (b) Deformation versus $De$ for various $Ca_E$, for $\rho_r=1$, $\mu_r=1$, $Re=1$, $\sigma_r=10$, $\epsilon_r=1.37$, $\beta_i = 1/9$. (from $PR_A^+$)}
    \label{Fig_deformation_PRA_plus}
\end{figure*}

\begin{figure*}[ht!]
    \centering
    \includegraphics{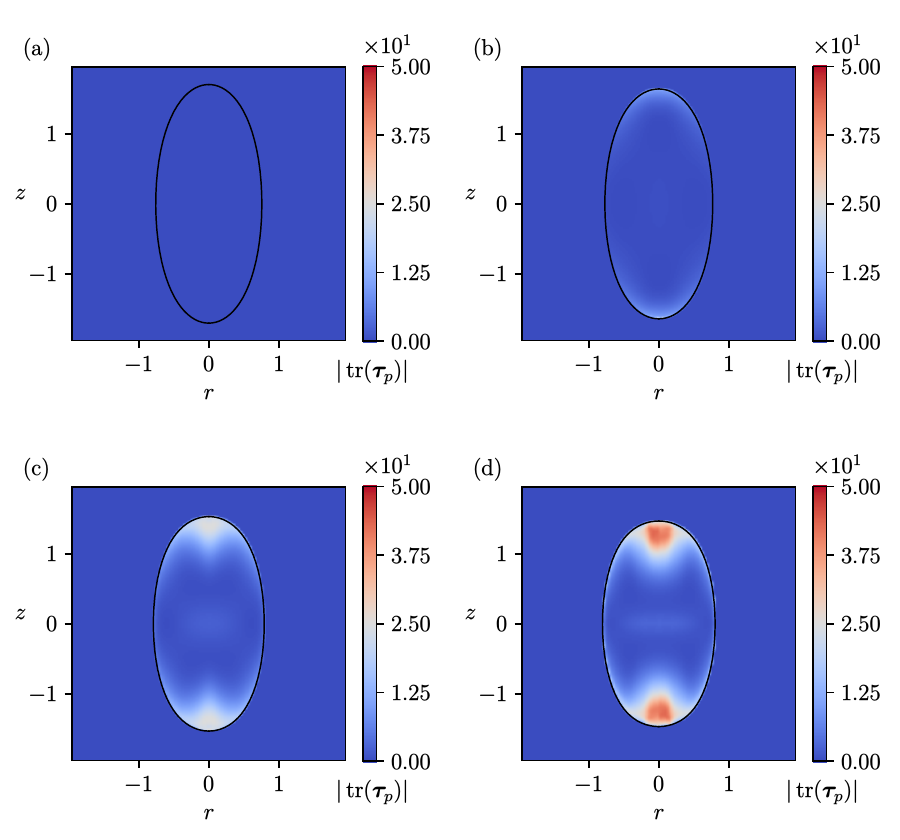}
    \caption{Contours of trace of polymeric stress for $Ca_E=0.36$, $\rho_r=1$, $\mu_r=1$, $Re=1$, $\sigma_r=10$, $\epsilon_r=1.37$, $\beta_i = 1/9$. (a) $De=0$, (b) $De=1$ (c) $De=5$, (d) $De=10$.}
    \label{Fig_taupTrace_PRA_plus}
\end{figure*}

The quantitative dependence of deformation on $Ca_E$ and $De$ is presented in \autoref{Fig_deformation_PRA_plus}.
As shown in \autoref{Fig_deformation_PRA_plus}(a), the deformation curves exhibit two distinct branches. The lower branch corresponds to spheroidal equilibrium shapes, valid up to the critical $Ca_E$, while the upper branch represents multi-lobed equilibrium states attained beyond the critical threshold. The lower branch consistently displays a positive curvature across all Deborah numbers.
Importantly, for a given $Ca_E$, the deformation magnitude decreases with increasing $De$, indicating the elasticity-induced suppression of interfacial deformation. This trend is reinforced in \autoref{Fig_deformation_PRA_plus}(b), where deformation plotted against $De$ shows a systematic reduction across all $Ca_E$ values.

To understand the reduction in deformation with increasing Deborah number, we plot the contours of the trace of the polymeric stress tensor for $De=0,\,1,\,5,$ and $10$ at a fixed capillary number $Ca_E=0.36$, as shown in \autoref{Fig_taupTrace_PRA_plus}. For $De=0$, the polymeric stress contribution is absent and the drop deformation is governed primarily by the balance between electric stresses and surface tension. As $De$ increases, the contours show a progressive increase in the magnitude and spatial extent of the polymeric stress inside the drop, with pronounced stress accumulation near the poles of the drop. This increase in polymeric stress corresponds to enhanced elastic stretching of polymer chains, which generates additional normal stresses that oppose the flow- and electric-field-induced deformation. As a result, the elastic resistance provided by the polymeric stresses suppresses interfacial stretching, leading to a monotonic decrease in drop deformation with increasing $De$.

\subsubsection{\texorpdfstring{$(\sigma_r, \epsilon_r)$ from $PR_B^+$ region}{Lg}}
\label{Sec_PRB_plus}
\begin{figure*}[ht!]
    \centering
    \includegraphics{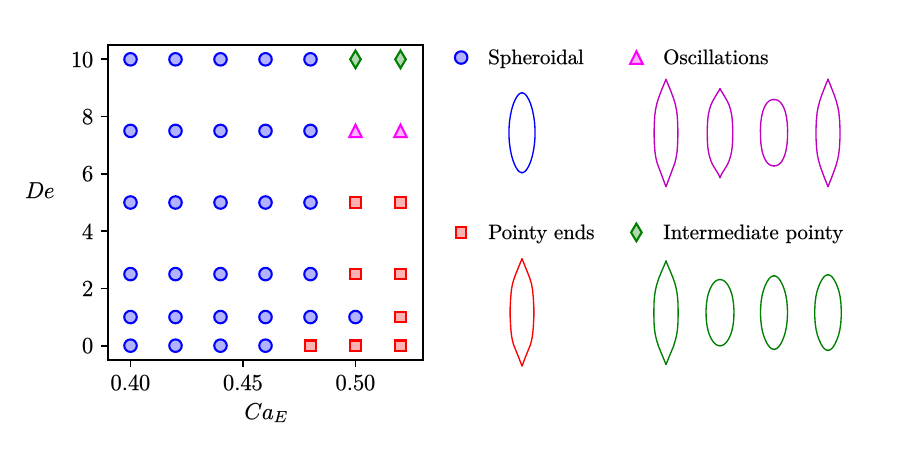}
    \caption{Phase plot on $(Ca_E, De)$ plane representing the existence of spheroidal shapes, shapes with pointed ends, oscillations of drop between spheroidal and pointy shapes and intermediate existence of pointy shapes before achieving stable spheroidal shapes}
    \label{Fig_phase_plot_PRB_plus}
\end{figure*}
An Oldroyd-B drop with property ratios 
$(\sigma_r, \epsilon_r)$ lying in the 
$PR_B^+$ regime deforms into a stable spheroidal shape under a steady electric field, provided the electric capillary number remains below a critical threshold 
($Ca_E^{crit}$). Beyond this critical value, however, the deformation dynamics becomes markedly more complex. The drop may elongate into a configuration with pointed ends, oscillate between spheroidal and pointed-end shapes, or even form transient pointed ends before eventually relaxing back to a stable spheroidal equilibrium.
This range of deformation behaviors is mapped in the phase diagram shown in \autoref{Fig_phase_plot_PRB_plus}, constructed in the 
($Ca_E$,$De$) plane. Blue markers denote the conditions yielding stable spheroidal shapes, whereas red markers correspond to cases that evolve into persistent pointed-end configurations. Magenta markers identify regimes where the interface oscillates between spheroidal and pointed shapes, while green markers capture scenarios in which pointed-end shapes arise transiently before the drop settles into a spheroidal state.

\begin{figure*}[ht!]
    \centering
    \includegraphics{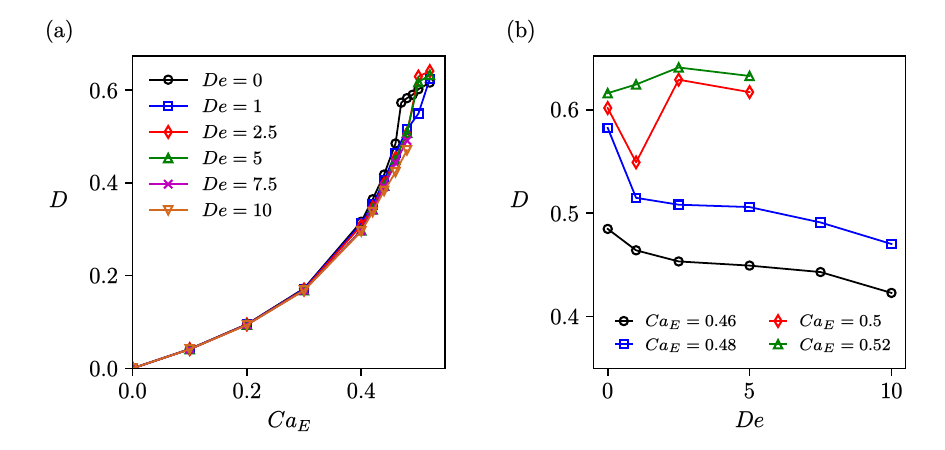}
    \caption{(a) Deformation versus $Ca_E$ for various $De$, (b) Deformation versus $De$ for various $Ca_E$, for $\rho_r=1$, $\mu_r=1$, $Re=1$, $\sigma_r=25$, $\epsilon_r=50$, $\beta_i = 1/9$. (from $PR_B^+$)}
    \label{Fig_deformation_PRB_plus_1}
\end{figure*}
\begin{figure*}[ht!]
    \centering
    \includegraphics{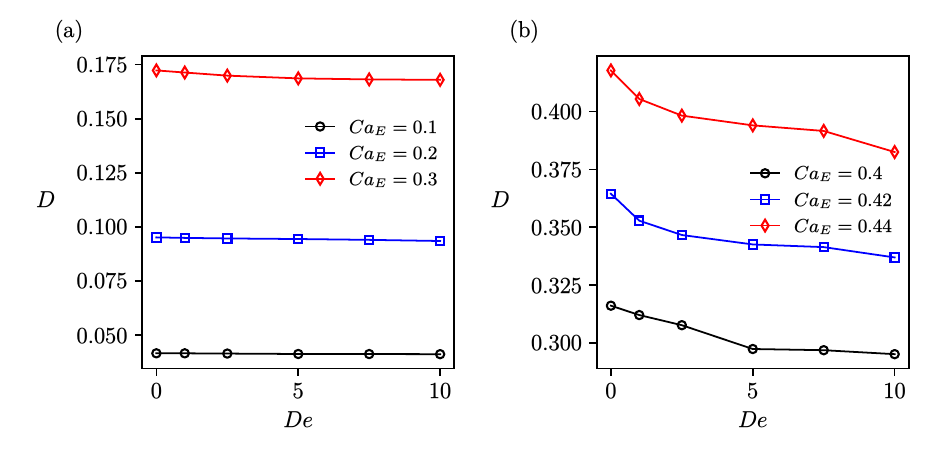}
    \caption{(a) Deformation versus $De$ for various $Ca_E$ for lower electric capillary numbers, (b) Deformation versus $De$ for various $Ca_E$ for intermediate electric capillary numbers, for $\rho_r=1$, $\mu_r=1$, $Re=1$, $\sigma_r=25$, $\epsilon_r=50$, $\beta_i = 1/9$. (from $PR_B^+$)}
    \label{Fig_deformation_PRB_plus_2}
\end{figure*}

\begin{figure*}[ht!]
    \centering
    \includegraphics{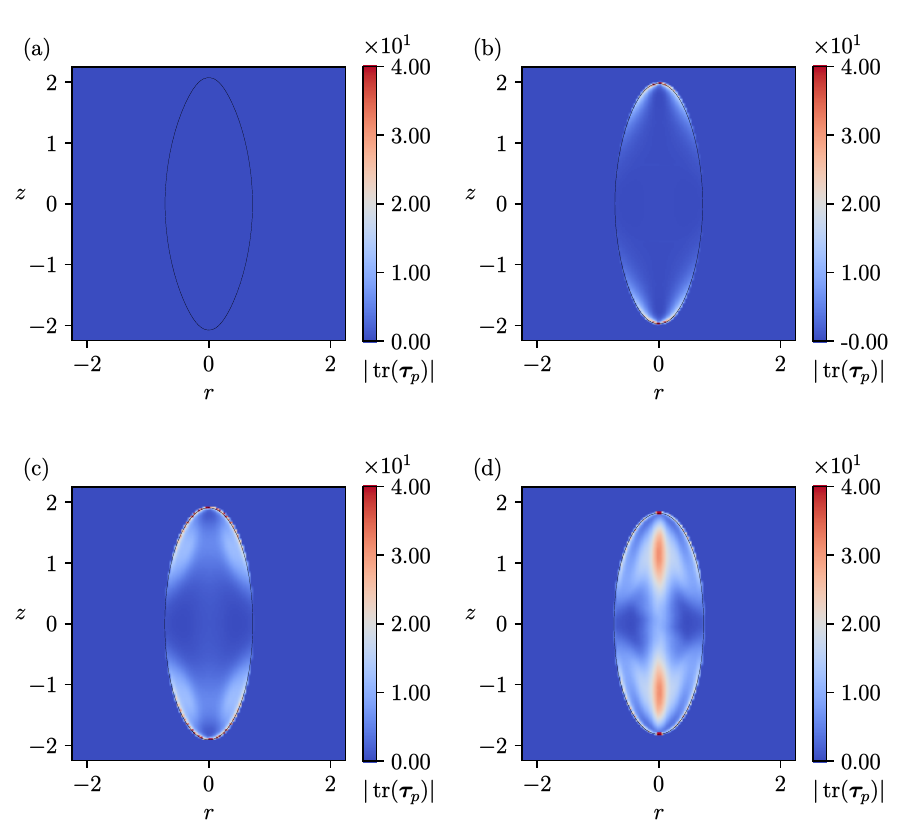}
    \caption{Contours of trace of polymeric stress for $Ca_E=0.46$, $\rho_r=1$, $\mu_r=1$, $Re=1$, $\sigma_r=25$, $\epsilon_r=50$, $\beta_i = 1/9$. (a) $De=0$, (b) $De=1$ (c) $De=5$, (d) $De=10$. The values of $|\trace(\bm{\tau}_p)|$ are significantly higher at the poles of the drop (of the order of 100 for $De = 1$ and 5, and of the order of 500 for $De = 10$). The colorbar limits are adjusted to clearly capture the internal field variations.}
    \label{Fig_taupTrace_PRB_plus}
\end{figure*}

\begin{figure*}[ht!]
    \centering
    \includegraphics{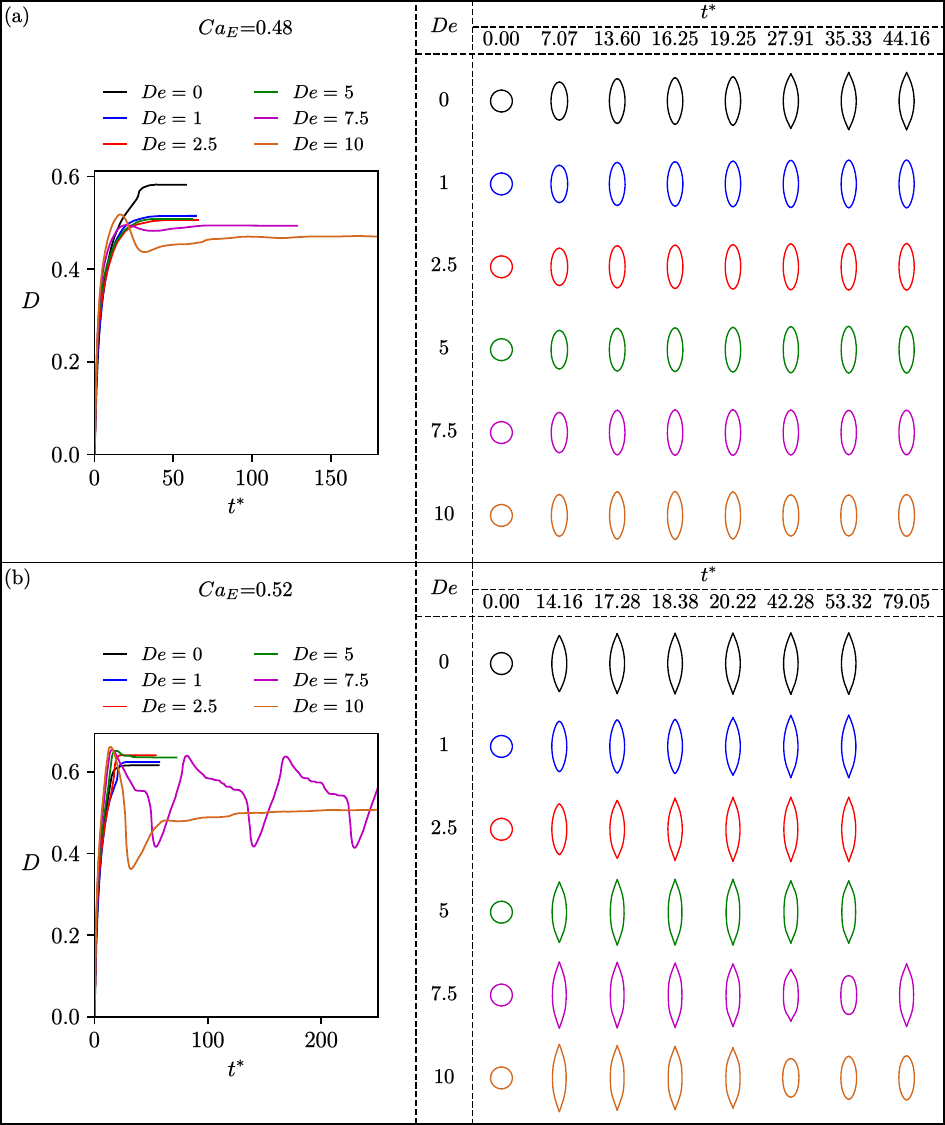}
    \caption{Variation in deformation parameter with time, and temporal evolution of the drop interface for various $De$ for $\rho_r=1$, $\mu_r=1$, $Re=1$, $\sigma_r=25$, $\epsilon_r=50$, $\beta_i = 1/9$. (a) $Ca_E=0.48$ (b) $Ca_E=0.52$.}
    \label{Fig_timeEvolution_PRB_plus}
\end{figure*}

\autoref{Fig_deformation_PRB_plus_1} and \autoref{Fig_deformation_PRB_plus_2} show the variation in the deformation parameter with electric capillary number ($Ca_E$) and Deborah number ($De$).
\autoref{Fig_deformation_PRB_plus_1}(a) shows the dependence of drop deformation on $Ca_E$ across different $De$ values. For all $De$, deformation increases monotonically with $Ca_E$, and the concave upward curvature indicates an accelerated rise in the deformation at higher $Ca_E$. The curves for $De=7.5$ and $10$ are truncated at $Ca_E=0.48$, as for $Ca_E \geq 0.5$ the drop either oscillates between spheroidal and pointed-end shapes or forms transient pointed ends 
% {\bf can we put a supplementary video for this?}.
\autoref{Fig_deformation_PRB_plus_1}(b) illustrates the variation of deformation with $De$ for different $Ca_E$. At $Ca_E=0.46$, the drop maintains a stable spheroidal shape across all $De$, with deformation decreasing monotonically as $De$ increases. At $Ca_E=0.48$, pointed-end deformation occurs only for $De=0$, while for $De \geq 1$ the drop remains spheroidal with deformation decreases with $De$. For $Ca_E=0.5$, pointed ends arise for $De=0, 2.5,$ and $5$, while for $De=7.5$ and $10$ oscillatory or transiently pointed shapes are observed; notably, a spheroidal equilibrium is preserved at $De=1$. Interestingly, the deformation parameter $D$ initially decreases with increase in $De$ and then increases. When $Ca_E$ is further increased to $0.52$, the drop develops pointed ends for all $De \leq 5$, with deformation increasing with $De$. However, for $De=7.5$ and $10$, the drop either oscillates between spheroidal and pointed configurations or evolves into a spheroidal shape after transient pointed deformation.
The sensitivity of drop deformation to $De$ at low and moderate $Ca_E$ is examined in \autoref{Fig_deformation_PRB_plus_2}. At low $Ca_E$, \autoref{Fig_deformation_PRB_plus_2}(a) shows that deformation is only weakly dependent on $De$, with a slight decrease as $De$ increases. In contrast, at moderate $Ca_E$, \autoref{Fig_deformation_PRB_plus_2}(b) demonstrates a consistent monotonic decrease in the deformation with increasing $De$.

The effect of increasing Deborah number on drop deformation is analyzed through contours of the trace of the polymeric stress tensor for $De=0\,,1\,,5,$ and $10$ at $Ca_E=0.46$, as shown in \autoref{Fig_taupTrace_PRB_plus}. For $De=0$, polymeric stresses are absent and the drop deformation is governed by the balance between electric stresses and surface tension. With increasing $De$, the magnitude of the polymeric stress increases significantly, as evidenced by the rising values of $\trace(\bm{\tau}_p)$ in the contours. Unlike the $PR_A^+$ case, the polymeric stresses in the $PR_B^+$ are strongly localized near the drop poles. This accumulation of polymeric stress at the poles generates large elastic normal stresses that oppose the electric tensile stresses responsible for drop elongation. As $De$ increases, this elastic resistance at the poles becomes stronger, resisting the axial stretching of the interface and thereby reducing the overall drop deformation.

\autoref{Fig_timeEvolution_PRB_plus} provides further insight through the temporal evolution of deformation and interfacial dynamics at $Ca_E = 0.48$ and $Ca_E = 0.52$ for various $De$. At $Ca_E=0.48$, pointed ends occur only for $De=0$, while for $De \geq 1$ the drop consistently evolves into a stable spheroidal shape. The time histories of deformation show that drops with $De \leq 5$ reach steady state rapidly, whereas for $De=7.5$ and $10$, the approach to equilibrium is slower, characterized by an overshoot in deformation before relaxing to a lower steady value. The overshoot is especially pronounced at $De=10$.
At the higher $Ca_E=0.52$, pointed-end shapes are observed across all $De$. For $De \leq 5$, the drop settles into a pointed steady state. For $De=7.5$, the drop oscillates between spheroidal and pointed configurations, while for $De=10$, deformation initially overshoots to a pointed shape before reducing to a spheroidal equilibrium.

\subsubsection{\texorpdfstring{$(\sigma_r, \epsilon_r)$ from $OB^-$ region}{Lg}}
\label{Sec_OB_plus}
\begin{figure*}[ht!]
    \centering
    \includegraphics{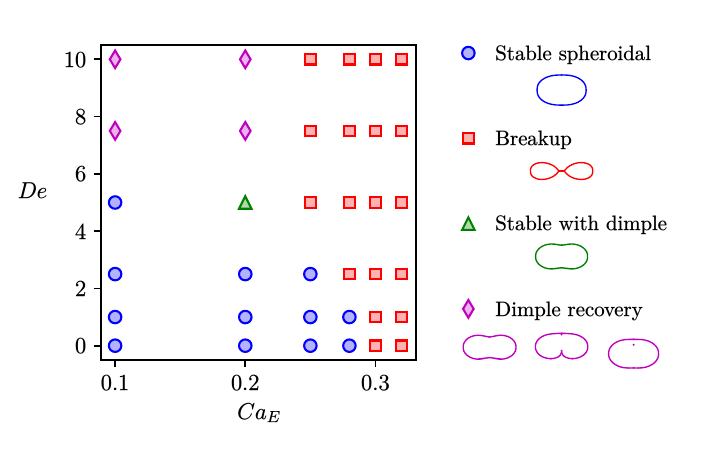}
    \caption{Phase plot on $(Ca_E, De)$ plane representing the existence of stable spheroidal shapes, stable shape with dimple, drop breakup and intermediate dimple formation with subsequent position oscillations.}
    \label{Fig_phase_plot_OB_minus}
\end{figure*}

\begin{figure*}
	\centering
	\includegraphics{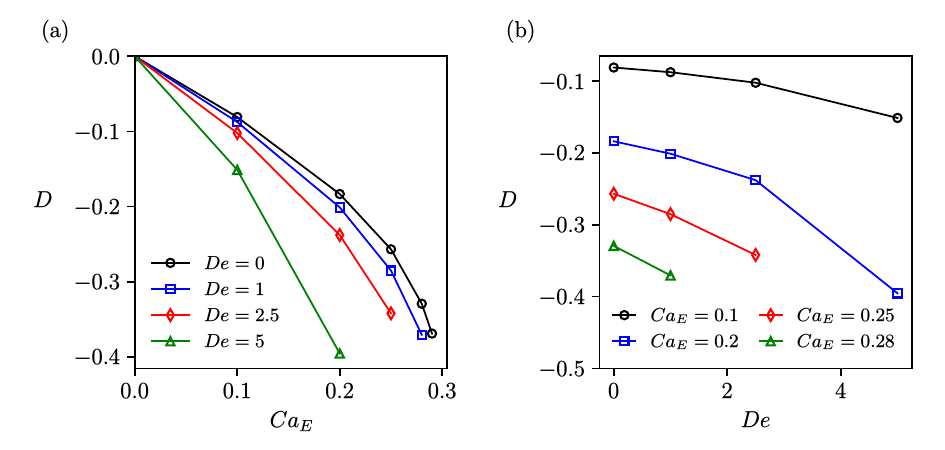}
	\caption{(a) Deformation versus $Ca_E$ for various $De$, (b) Deformation versus $De$ for various $Ca_E$, for $\rho_r=1$, $\mu_r=1$, $Re=1$, $\sigma_r=0.1$, $\epsilon_r=2$, $\beta_i = 1/9$. (from $OB^-$)}
	\label{Fig_deformation_OB_minus}
\end{figure*}

For an Oldroyd-B fluid drop characterized by $(\sigma_r, \epsilon_r)$ within the $OB^-$ regime, application of a steady external electric field induces primarily oblate spheroidal deformation. At electric capillary numbers below a critical threshold, the drop can also adopt stable configurations featuring dimples. When $Ca_E$ exceeds this critical value, the drop may destabilize, leading to breakup, or transiently form a dimple before recovering and undergoing longitudinal oscillations.
\autoref{Fig_phase_plot_OB_minus} presents a phase diagram in the ($Ca_E$, $De$) plane, capturing these varied deformation behaviors. In this diagram, blue markers correspond to stable oblate spheroidal shapes, red markers indicate drop breakup, green markers denote stable oblate shapes with dimples, and magenta markers represent cases where the drop temporarily attempts breakup but ultimately recovers and exhibits positional oscillations.

\autoref{Fig_deformation_OB_minus} shows the dependence of the drop deformation parameter on the electric capillary number ($Ca_E$) and the Deborah number ($De$), highlighting the conditions that lead to stable deformed shapes. In \autoref{Fig_deformation_OB_minus}(a), the deformation is seen to decrease (i.e., the magnitude of oblate deformation increases) with increasing $Ca_E$, indicating that stronger electric fields enhance the oblate nature of the drop. \autoref{Fig_deformation_OB_minus}(b) presents the variation in deformation with $De$ for several $Ca_E$ values. The results demonstrate that increasing the viscoelasticity of the drop amplifies the magnitude of oblate deformation.

\begin{figure*}[ht!]
    \centering
    \includegraphics{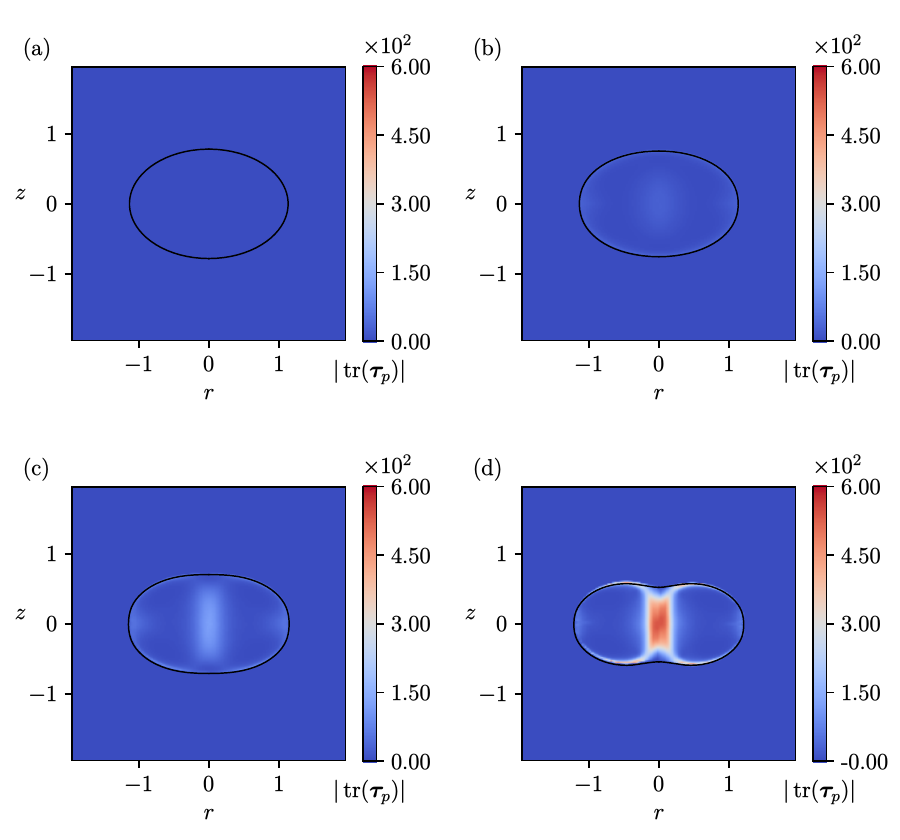}
    \caption{Contours of trace of polymeric stress for $Ca_E=0.36$, $\rho_r=1$, $\mu_r=1$, $Re=1$, $\sigma_r=10$, $\epsilon_r=1.37$, $\beta_i = 1/9$. (a) $De=0$, (b) $De=1$ (c) $De=2.5$, (d) $De=5$.}
    \label{Fig_taupTrace_OB_minus}
\end{figure*}

\begin{figure*}
	\centering
	\includegraphics{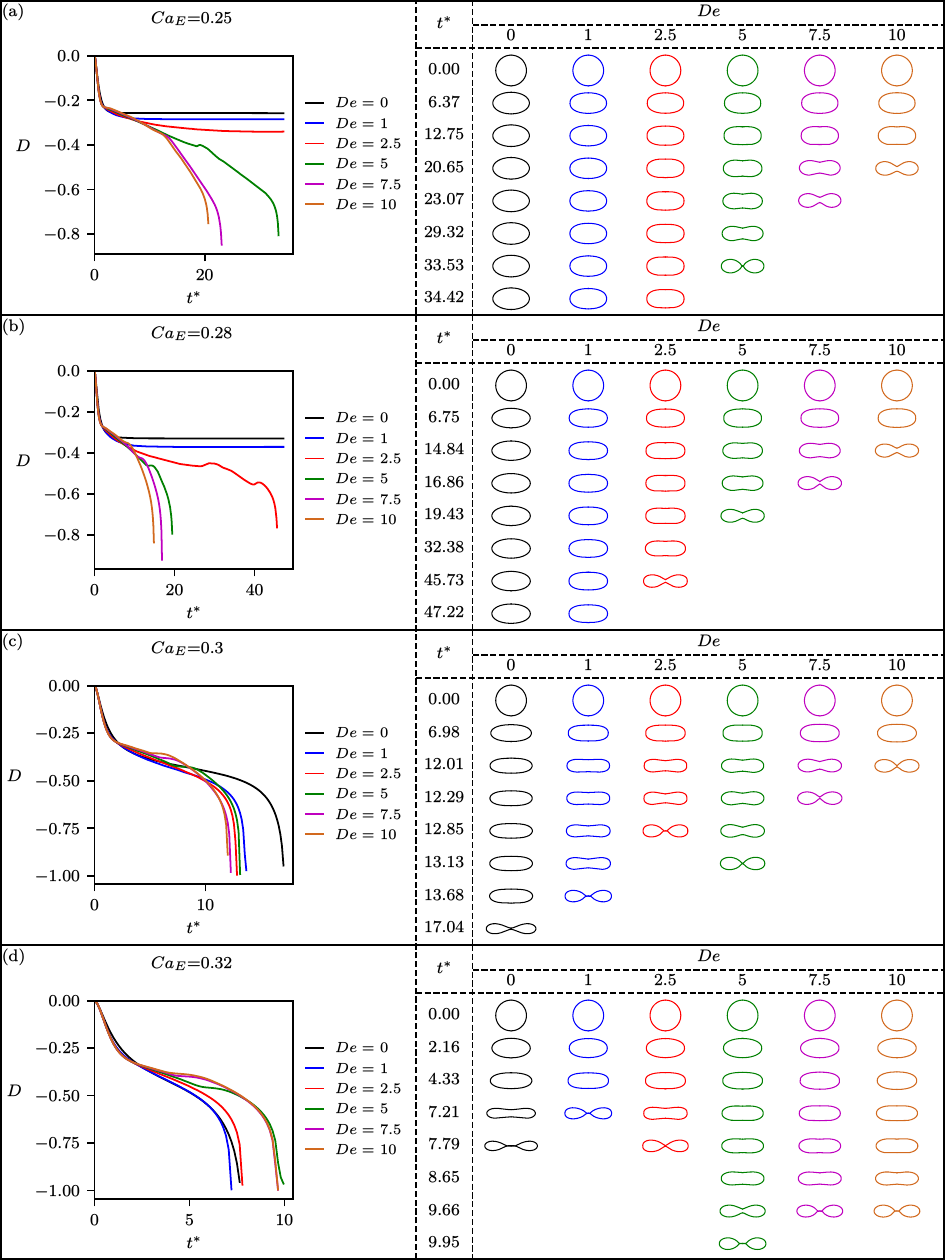}
	\caption{The variation of drop deformation with time and the evolution of the drop interface for various Deborah numbers for $\rho_r=1$, $\mu_r=1$, $Re=1$, $\sigma_r=0.1$, $\epsilon_r=2$, $\beta_i = 1/9$. (a) $Ca_E=0.25$ (b) $Ca_E=0.28$ (c) $Ca_E=0.3$ (d) $Ca_E=0.32$}
	\label{Fig_timeEvolution_OB_minus}
\end{figure*} 

The deformation behavior in the oblate regime is examined, using the contours of the trace of the polymeric stress tensor for $De=0,\,1,\,2.5,$ and $5$ at a fixed electric capillary number $Ca_E=0.2$, as shown in \autoref{Fig_taupTrace_OB_minus}. In this regime, the drop undergoes oblate deformation, characterized by negative values of the deformation parameter. For $De=0$, the deformation is governed solely by the balance between electric stresses and surface tension. As $De$ increases, the magnitude of the polymeric stress increases throughout the drop, as evidenced by the growing values of $\trace(\bm{\tau}_p)$. The increasing polymeric stresses with $De$ enhance the resistance to axial extension while promoting radial spreading, leading to an increase in the magnitude of negative deformation. This explains why, in contrast to the prolate cases, the oblate deformation becomes stronger with increasing $De$.

\autoref{Fig_timeEvolution_OB_minus} depicts the temporal evolution of the deformation and corresponding interface shapes for an Oldroyd-B drop with $(\sigma_r, \epsilon_r) = (0.1, 2)$ within the $OB^-$ region, at four electric capillary numbers: $Ca_E = 0.25, 0.28, 0.3,$ and $0.32$, and for varying Deborah numbers. The deformation versus dimensionless time curves are truncated at breakup, characterized by a sharp drop in deformation immediately preceding the event. At $Ca_E = 0.25$, the drop achieves a stable oblate steady-state for $De \leq 2.5$, whereas higher Deborah numbers ($De \geq 5$) result in breakup. For $Ca_E = 0.28$, stable configurations occur only for $De \leq 1$, with breakup for $De \geq 2.5$. At the higher $Ca_E$ values of $0.3$ and $0.32$, the drop fails to reach a stable shape across all examined Deborah numbers, invariably undergoing breakup.

\begin{figure*}[ht!]
    \centering
    \includegraphics{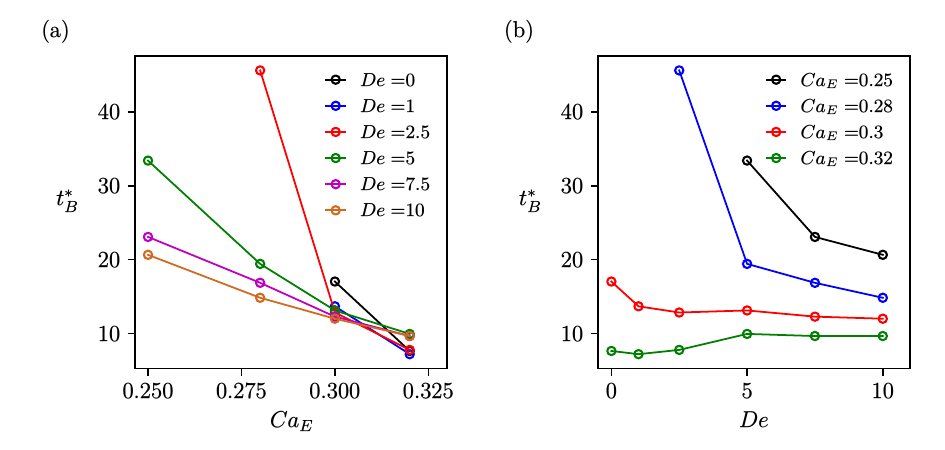}
    \caption{The variation in breakup time with (a) $Ca_E$ and (b) $De$ for $\rho_r=1$, $\mu_r=1$, $Re=1$, $\sigma_r=0.1$, $\epsilon_r=2$, $\beta_i = 1/9$.}
    \label{Fig_breakTimes_OB_minus}
\end{figure*}
\autoref{Fig_breakTimes_OB_minus}(a) and (b) present the variation in the dimensionless drop breakup time with electric capillary number ($Ca_E$) and Deborah number ($De$) for an Oldroyd-B fluid drop, directly reflecting the temporal evolution depicted in \autoref{Fig_timeEvolution_OB_minus}.
As shown in \autoref{Fig_breakTimes_OB_minus}(a), the dimensionless breakup time decreases with increasing $Ca_E$ for all the examined $De$, indicating that stronger electric fields accelerate drop breakup. Drop breakup is influenced by both $De$ and $Ca_E$: at low $De$, breakup occurs only at higher $Ca_E$ (e.g., for $De=0$ and $De=1$, breakup is observed for $Ca_E \geq 0.3$). For $De=2.5$, breakup initiates at $Ca_E \geq 0.29$, while for $De \geq 5$, breakup occurs at all $Ca_E \geq 0.25$.
\autoref{Fig_breakTimes_OB_minus}(b) shows the effect of $De$ on breakup time for different $Ca_E$. For $Ca_E = 0.25$ and $0.28$, the breakup time decreases monotonically with $De$, indicating that increasing viscoelasticity accelerates breakup. In contrast, for higher $Ca_E$ values ($Ca_E \geq 0.3$), the breakup time exhibits non-monotonic behavior with $De$. At $Ca_E = 0.3$, breakup times are nearly constant for $De \geq 1$, with a slightly higher value for $De=0$. Specifically, breakup times decrease with $De$ up to $De=2.5$, increase slightly at $De=5$, and then decrease again for $De$ approaching 10. For $Ca_E = 0.32$, breakup times for $De \geq 5$ cluster together and are higher than those for $De \leq 2.5$, which form a separate cluster.
At higher $De$, the rate of decrease of breakup time with increasing $Ca_E$ is slightly lower than at lower $De$. Consequently, at the maximum $Ca_E = 0.32$, breakup times across different Deborah numbers converge to similar values, with slightly shorter times for higher $De \geq 5$ compared to $De \leq 2.5$. Despite the non-monotonic dependence of breakup time on $De$ at high $Ca_E$, the differences in dimensionless breakup time across $De$ remain relatively small. Overall, increasing the elasticity of an Oldroyd-B drop generally enhances deformation and accelerates breakup, with only a minor delay observed at higher $De$ ($De \geq 5$) for $Ca_E = 0.32$.

\section{Conclusion}
In this study, we investigated the dynamics of an Oldroyd-B drop under an external electric field using numerical simulations. A representative pair of conductivity ratio ($\sigma_r$) and permittivity ratio ($\epsilon_r$) is selected from each of the six regions ($PR_A^+$, $PR_B^+$, $PR_A^-$, $PR_B^-$, $OB^+$, and $OB^-$) of the $\sigma_r$–$\epsilon_r$ phase space. In regions where both the first- and second-order deformation coefficients have the same sign ($PR_A^-$, $PR_B^-$, and $OB^+$), deviations from Newtonian behavior are negligible. 
In the $PR_A^+$ region, the drop develops multi-lobed shapes above a critical electric capillary number. In this region, elasticity reduces the steady state deformation and the critical $Ca_E$ increases with the Deborah number ($De$). 
For $PR_B^+$, the steady-state deformation decreases with $De$, while the critical $Ca_E$ exhibits non-monotonic behavior. At higher $De$, the deformation–time curve exhibits transient maxima and minima before reaching steady state, and in some cases, the drop undergoes oscillations between spheroidal and pointed shapes. 
In the $OB^-$ region, deformation magnitude increases with $De$, and the critical $Ca_E$ decreases. At low $Ca_E$ and high $De$, drops exhibit dimpling, attempt breakup, but recover and perform positional oscillations.

%% The Appendices part is started with the command \appendix;
%% appendix sections are then done as normal sections
\appendix
\section{Asymptotic solution for drop deformation} \label{appA}
The drop deformation obtained from asymptotic solution is given by,
\begin{widetext}
\begin{equation}
    D = Ca_E D^{(Ca_E)} + Ca_E^2 D^{(Ca_E^2)} + Ca_E De D^{(Ca_E De)}
\end{equation}
where,
\begin{equation}
    D^{(Ca_E)} = \frac{9}{80} \frac{1}{(1 + \mu_r) (2 + \sigma_r)^2} \Big[ (6 + 9 \mu_r) \sigma_r -(16 + 19\mu_r)\epsilon_r + 5 (1 + \mu_r) (1 + \sigma_r^2) \Big]
\end{equation}
\begin{equation}
\begin{split}
    D^{(Ca_E^2)} &= \frac{D^{(Ca_E)}}{(2 + \sigma_r)^3(1 + \mu_r)^2} \Bigg[\frac{(2743 + \mu_r (20616 + 15281 \mu_r))\sigma_r}{8400}\\
    &+ \frac{(5120 + \mu_r (31413 + 24997 \mu_r)) \sigma_r^2}{8400} + \frac{1}{80} (1 + \mu_r)^2 (-154 + 139 \sigma_r^3)\\
    &+ \epsilon_r \bigg\{ \frac{22096 + \mu_r (36627 + 15827 \mu_r)}{4200}\\
    &+ \sigma_r \left(\frac{-50480 - \mu_r(122133 + 70357\mu_r)}{8400}\right) \bigg\} \Bigg]
\end{split}
\end{equation}
\begin{equation}
\begin{split}
    D^{(Ca_E De)} = \frac{27(\epsilon_r - \sigma_r)^2}{56000 (1 + \mu_r)^3 (2 + \sigma_r)^4}
    \Big[ &\Lambda_e (1 - \beta_e)(408 - 443\mu_r)\\
    &- 4\Lambda_i \mu_r (1 - \beta_i)(787 + 998 \mu_r) \Big] \; .
    \label{eqn:D_CDe_full}
\end{split}
\end{equation}
For an Oldroyd-B drop suspended in a Newtonian ambient, $\beta_e = 1$ and $\Lambda_e = 0$. Thus,
\begin{equation}
    D^{(Ca_E De)} = -\frac{27(\epsilon_r - \sigma_r)^2}{14000 (1 + \mu_r)^3 (2 + \sigma_r)^4}
    \Big(\Lambda_i \mu_r (1 - \beta_i)(787 + 998 \mu_r) \Big) \; .
    \label{eqn:D_CDe_ON_full}
\end{equation}
\end{widetext}
For detailed derivation of the above expressions refer to \citet{DAS2026105633}.
\bibliography{references}

@article{basaran2013nonstandard,
  title={Nonstandard inkjets},
  author={Basaran, Osman A and Gao, Haijing and Bhat, Pradeep P},
  journal={Annual Review of Fluid Mechanics},
  volume={45},
  number={1},
  pages={85--113},
  year={2013},
  publisher={Annual Reviews}
}

@article{lau2017ink,
  title={Ink-jet printing of micro-electro-mechanical systems (MEMS)},
  author={Lau, Gih-Keong and Shrestha, Milan},
  journal={Micromachines},
  volume={8},
  number={6},
  pages={194},
  year={2017},
  publisher={MDPI}
}

@article{eow2002electrostatic,
  title={Electrostatic enhancement of coalescence of water droplets in oil: a review of the technology},
  author={Eow, John S and Ghadiri, Mojtaba},
  journal={Chemical Engineering Journal},
  volume={85},
  number={2-3},
  pages={357--368},
  year={2002},
  publisher={Elsevier}
}

@article{alvarado2010enhanced,
  title={Enhanced oil recovery: an update review},
  author={Alvarado, Vladimir and Manrique, Eduardo},
  journal={Energies},
  volume={3},
  number={9},
  pages={1529--1575},
  year={2010},
  publisher={MDPI}
}

@article{zhang2011application,
  title={Application of variable frequency technique on electrical dehydration of water-in-oil emulsion},
  author={Zhang, Yanzhen and Liu, Yonghong and Ji, Renjie and Wang, Fei and Cai, Baoping and Li, Hang},
  journal={Colloids and Surfaces A: Physicochemical and Engineering Aspects},
  volume={386},
  number={1-3},
  pages={185--190},
  year={2011},
  publisher={Elsevier}
}

@article{kelly1984electrostatic,
  title={The electrostatic atomization of hydrocarbons},
  author={Kelly, AJ},
  journal={J. Inst. Energy;(United Kingdom)},
  volume={57},
  number={431},
  year={1984}
}

@incollection{law2018electrostatic,
  title={Electrostatic atomization and spraying},
  author={Law, S Edward},
  booktitle={Handbook of electrostatic processes},
  pages={429--456},
  year={2018},
  publisher={CRC Press}
}

@article{hines1966electrostatic,
  title={Electrostatic atomization and spray painting},
  author={Hines, RL},
  journal={Journal of Applied Physics},
  volume={37},
  number={7},
  pages={2730--2736},
  year={1966},
  publisher={American Institute of Physics}
}

@article{moreau2015electrohydrodynamic,
  title={Electrohydrodynamic force produced by a corona discharge between a wire active electrode and several cylinder electrodes--Application to electric propulsion},
  author={Moreau, Eric and Benard, Nicolas and Alicalapa, Fr{\'e}d{\'e}ric and Douy{\`e}re, Alexandre},
  journal={Journal of Electrostatics},
  volume={76},
  pages={194--200},
  year={2015},
  publisher={Elsevier}
}

@inproceedings{huh2019numerical,
  title={Numerical simulation of electrospray thruster extraction},
  author={Huh, Henry and Wirz, Richard E},
  booktitle={Proceedings of the 36th International Electric Propulsion Conference, Vienna, Austria},
  pages={15--20},
  year={2019}
}

@article{laser2004review,
  title={A review of micropumps},
  author={Laser, Daniel J and Santiago, Juan G},
  journal={Journal of micromechanics and microengineering},
  volume={14},
  number={6},
  pages={R35},
  year={2004},
  publisher={IOP Publishing}
}

@article{stone2004engineering,
  title={Engineering flows in small devices: microfluidics toward a lab-on-a-chip},
  author={Stone, Howard A and Stroock, Abraham D and Ajdari, Armand},
  journal={Annu. Rev. Fluid Mech.},
  volume={36},
  number={1},
  pages={381--411},
  year={2004},
  publisher={Annual Reviews}
}

@article{phan2025demand,
  title={On-demand electrostatic droplet sorting and splitting},
  author={Phan, Hoang Anh and Nguyen, Kien and Pham, Phong Tuan and Do Quang, Loc and Thu, Hang Bui and Lam, Dang Bao and Jen, Chun-Ping and Thanh, Tung Bui and Duc, Trinh Chu},
  journal={Sensors and Actuators A: Physical},
  volume={385},
  pages={116311},
  year={2025},
  publisher={Elsevier}
}

@article{wilson1921iii,
  title={III. Investigations on lighting discharges and on the electric field of thunderstorms},
  author={Wilson, Charles Thomson Rees},
  journal={Philosophical Transactions of the Royal Society of London. Series A, Containing Papers of a Mathematical or Physical Character},
  volume={221},
  number={582-593},
  pages={73--115},
  year={1921},
  publisher={The Royal Society London}
}

@article{blanchard1963electrification,
  title={The electrification of the atmosphere by particles from bubbles in the sea},
  author={Blanchard, Duncan C},
  journal={Progress in oceanography},
  volume={1},
  pages={73--202},
  year={1963},
  publisher={Elsevier}
}

@article{simpson1909electricity,
  title={On the electricity of rain and its origin in thunderstorm},
  author={Simpson, GC},
  journal={Phil. Trans., A},
  volume={209},
  pages={397--413},
  year={1909}
}

@article{o1953distortion,
  title={The distortion of aerosol droplets by an electric field},
  author={O'Konski, Chester T and Thacher Jr, Henry C},
  journal={The Journal of Physical Chemistry},
  volume={57},
  number={9},
  pages={955--958},
  year={1953},
  publisher={ACS Publications}
}

@article{taylor1964disintegration,
  title={Disintegration of water drops in an electric field},
  author={Taylor, Geoffrey Ingram},
  journal={Proceedings of the Royal Society of London. Series A. Mathematical and Physical Sciences},
  volume={280},
  number={1382},
  pages={383--397},
  year={1964},
  publisher={The Royal Society London}
}

@article{allan1962particle,
  title={Particle behaviour in shear and electric fields I. Deformation and burst of fluid drops},
  author={Allan, RS and Mason, SG},
  journal={Proceedings of the Royal Society of London. Series A. Mathematical and Physical Sciences},
  volume={267},
  number={1328},
  pages={45--61},
  year={1962},
  publisher={The Royal Society London}
}

@article{o1957electric,
  title={Electric free energy and the deformation of droplets in electrically conducting systems},
  author={O'Konski, Chester T and Harris, Frank E},
  journal={The Journal of Physical Chemistry},
  volume={61},
  number={9},
  pages={1172--1174},
  year={1957},
  publisher={ACS Publications}
}

@article{taylor1966studies,
  title={Studies in electrohydrodynamics. I. The circulation produced in a drop by an electric field},
  author={Taylor, Geoffrey Ingram},
  journal={Proceedings of the Royal Society of London. Series A. Mathematical and Physical Sciences},
  volume={291},
  number={1425},
  pages={159--166},
  year={1966},
  publisher={The Royal Society London}
}

@article{melcher1969electrohydrodynamics,
  title={Electrohydrodynamics: a review of the role of interfacial shear stresses},
  author={Melcher, JR and Taylor, GI},
  journal={Annual review of fluid mechanics},
  volume={1},
  number={1},
  pages={111--146},
  year={1969},
  publisher={Annual Reviews 4139 El Camino Way, PO Box 10139, Palo Alto, CA 94303-0139, USA}
}

@article{torza1971electrohydrodynamic,
  title={Electrohydrodynamic deformation and bursts of liquid drops},
  author={Torza, S and Cox, RG and Mason, SG},
  journal={Philosophical Transactions of the Royal Society of London. Series A, Mathematical and Physical Sciences},
  volume={269},
  number={1198},
  pages={295--319},
  year={1971},
  publisher={The Royal Society London}
}

@article{ajayi1978note,
  title={A note on Taylor’s electrohydrodynamic theory},
  author={Ajayi, OO},
  journal={Proceedings of the Royal Society of London. A. Mathematical and Physical Sciences},
  volume={364},
  number={1719},
  pages={499--507},
  year={1978},
  publisher={The Royal Society London}
}

@article{saville1997electrohydrodynamics,
  title={Electrohydrodynamics: the Taylor-Melcher leaky dielectric model},
  author={Saville, DA1435033},
  journal={Annual review of fluid mechanics},
  volume={29},
  number={1},
  pages={27--64},
  year={1997},
  publisher={Annual Reviews 4139 El Camino Way, PO Box 10139, Palo Alto, CA 94303-0139, USA}
}

@article{ha1995effects,
  title={Effects of surfactant on the deformation and stability of a drop in a viscous fluid in an electric field},
  author={Ha, Jong-Wook and Yang, Seung-Man},
  journal={Journal of colloid and interface science},
  volume={175},
  number={2},
  pages={369--385},
  year={1995},
  publisher={Elsevier}
}

@article{bentenitis2005droplet,
  title={Droplet deformation in dc electric fields: the extended leaky dielectric model},
  author={Bentenitis, Nikolaos and Krause, Sonja},
  journal={Langmuir},
  volume={21},
  number={14},
  pages={6194--6209},
  year={2005},
  publisher={ACS Publications}
}

@article{zabarankin2013liquid,
  title={A liquid spheroidal drop in a viscous incompressible fluid under a steady electric field},
  author={Zabarankin, Michael},
  journal={Siam journal on applied mathematics},
  volume={73},
  number={2},
  pages={677--699},
  year={2013},
  publisher={SIAM}
}

@article{nganguia2013equilibrium,
  title={Equilibrium electro-deformation of a surfactant-laden viscous drop},
  author={Nganguia, Herve and Young, Y-N and Vlahovska, Petia M and Blawzdziewicz, Jerzy and Zhang, J and Lin, H},
  journal={Physics of Fluids},
  volume={25},
  number={9},
  year={2013},
  publisher={AIP Publishing}
}

@article{deshmukh2013deformation,
  title={Deformation and breakup of a leaky dielectric drop in a quadrupole electric field},
  author={Deshmukh, Shivraj D and Thaokar, Rochish M},
  journal={Journal of Fluid Mechanics},
  volume={731},
  pages={713--733},
  year={2013},
  publisher={Cambridge University Press}
}

@article{he2013electrorotation,
  title={Electrorotation of a viscous droplet in a uniform direct current electric field},
  author={He, Hui and Salipante, Paul F and Vlahovska, Petia M},
  journal={Physics of Fluids},
  volume={25},
  number={3},
  year={2013},
  publisher={AIP Publishing}
}

@article{yariv2016electrohydrodynamic,
  title={Electrohydrodynamic rotation of drops at large electric Reynolds numbers},
  author={Yariv, Ehud and Frankel, Itzchak},
  journal={Journal of Fluid Mechanics},
  volume={788},
  pages={R2},
  year={2016},
  publisher={Cambridge University Press}
}

@article{vlahovska2016electrohydrodynamic,
  title={Electrohydrodynamic instabilities of viscous drops},
  author={Vlahovska, Petia M},
  journal={Physical Review Fluids},
  volume={1},
  number={6},
  pages={060504},
  year={2016},
  publisher={APS}
}

@article{vlahovska2019electrohydrodynamics,
  title={Electrohydrodynamics of drops and vesicles},
  author={Vlahovska, Petia M},
  journal={Annual Review of Fluid Mechanics},
  volume={51},
  number={1},
  pages={305--330},
  year={2019},
  publisher={Annual Reviews}
}

@article{sengupta2017role,
  title={The role of surface charge convection in the electrohydrodynamics and breakup of prolate drops},
  author={Sengupta, Rajarshi and Walker, Lynn M and Khair, Aditya S},
  journal={Journal of Fluid Mechanics},
  volume={833},
  pages={29--53},
  year={2017},
  publisher={Cambridge University Press}
}

@article{sozou1972electrohydrodynamics,
  title={Electrohydrodynamics of a liquid drop: the time-dependent problem},
  author={Sozou, C},
  journal={Proceedings of the Royal Society of London. A. Mathematical and Physical Sciences},
  volume={331},
  number={1585},
  pages={263--272},
  year={1972},
  publisher={The Royal Society London}
}

@article{moriya1986deformation,
  title={Deformation of droplets suspended in viscous media in an electric field. 1. Rate of deformation},
  author={Moriya, Satoru and Adachi, Keiichiro and Kotaka, Tadao},
  journal={Langmuir},
  volume={2},
  number={2},
  pages={155--160},
  year={1986},
  publisher={ACS Publications}
}

@article{esmaeeli2011transient,
  title={Transient electrohydrodynamics of a liquid drop},
  author={Esmaeeli, Asghar and Sharifi, Payam},
  journal={Physical Review E—Statistical, Nonlinear, and Soft Matter Physics},
  volume={84},
  number={3},
  pages={036308},
  year={2011},
  publisher={APS}
}

@article{zhang2013transient,
  title={Transient solution for droplet deformation under electric fields},
  author={Zhang, Jia and Zahn, Jeffrey D and Lin, Hao},
  journal={Physical Review E—Statistical, Nonlinear, and Soft Matter Physics},
  volume={87},
  number={4},
  pages={043008},
  year={2013},
  publisher={APS}
}

@article{lanauze2013influence,
  title={The influence of inertia and charge relaxation on electrohydrodynamic drop deformation},
  author={Lanauze, Javier A and Walker, Lynn M and Khair, Aditya S},
  journal={Physics of Fluids},
  volume={25},
  number={11},
  year={2013},
  publisher={AIP Publishing}
}

@article{lanauze2015nonlinear,
  title={Nonlinear electrohydrodynamics of slightly deformed oblate drops},
  author={Lanauze, Javier A and Walker, Lynn M and Khair, Aditya S},
  journal={Journal of Fluid Mechanics},
  volume={774},
  pages={245--266},
  year={2015},
  publisher={Cambridge University Press}
}

@article{esmaeeli2020transient,
  title={Transient electrohydrodynamics of a liquid drop at finite Reynolds numbers},
  author={Esmaeeli, Asghar and Behjatian, Ali},
  journal={Journal of Fluid Mechanics},
  volume={893},
  pages={A26},
  year={2020},
  publisher={Cambridge University Press}
}

@article{nishiwaki1988deformation,
  title={Deformation of viscous droplets in an electric field: Poly (propylene oxide)/poly (dimethylsiloxane) systems},
  author={Nishiwaki, Tsuyoshi and Adachi, Keiichiro and Kotaka, Tadao},
  journal={Langmuir},
  volume={4},
  number={1},
  pages={170--175},
  year={1988},
  publisher={ACS Publications}
}

@article{vizika1992electrohydrodynamic,
  title={The electrohydrodynamic deformation of drops suspended in liquids in steady and oscillatory electric fields},
  author={Vizika, O and Saville, DA},
  journal={Journal of fluid Mechanics},
  volume={239},
  pages={1--21},
  year={1992},
  publisher={Cambridge University Press}
}

@article{karp2024electrohydrodynamic,
  title={Electrohydrodynamic flows inside a neutrally buoyant leaky dielectric drop},
  author={Karp, Joel R and Lecordier, Bertrand and Shadloo, Mostafa S},
  journal={Physics of Fluids},
  volume={36},
  number={5},
  year={2024},
  publisher={AIP Publishing}
}

@article{brosseau2017streaming,
  title={Streaming from the equator of a drop in an external electric field},
  author={Brosseau, Quentin and Vlahovska, Petia M},
  journal={Physical review letters},
  volume={119},
  number={3},
  pages={034501},
  year={2017},
  publisher={APS}
}

@article{miksis1981shape,
  title={Shape of a drop in an electric field},
  author={Miksis, Michael J},
  journal={The Physics of Fluids},
  volume={24},
  number={11},
  pages={1967--1972},
  year={1981},
  publisher={AIP Publishing}
}

@article{sherwood1988breakup,
  title={Breakup of fluid droplets in electric and magnetic fields},
  author={Sherwood, JD},
  journal={Journal of Fluid Mechanics},
  volume={188},
  pages={133--146},
  year={1988},
  publisher={Cambridge University Press}
}

@article{feng1996computational,
  title={A computational analysis of electrohydrodynamics of a leaky dielectric drop in an electric field},
  author={Feng, James Q and Scott, Timothy C},
  journal={Journal of Fluid Mechanics},
  volume={311},
  pages={289--326},
  year={1996},
  publisher={Cambridge University Press}
}

@article{feng1999electrohydrodynamic,
  title={Electrohydrodynamic behaviour of a drop subjected to a steady uniform electric field at finite electric Reynolds number},
  author={Feng, James Q},
  journal={Proceedings of the Royal Society of London. Series A: Mathematical, Physical and Engineering Sciences},
  volume={455},
  number={1986},
  pages={2245--2269},
  year={1999},
  publisher={The Royal Society}
}

@article{lac2007axisymmetric,
  title={Axisymmetric deformation and stability of a viscous drop in a steady electric field},
  author={Lac, Etienne and Homsy, GM},
  journal={Journal of Fluid Mechanics},
  volume={590},
  pages={239--264},
  year={2007},
  publisher={Cambridge University Press}
}

@article{dubash2007behaviour,
  title={Behaviour of a conducting drop in a highly viscous fluid subject to an electric field},
  author={Dubash, N and Mestel, AJ},
  journal={Journal of Fluid Mechanics},
  volume={581},
  pages={469--493},
  year={2007},
  publisher={Cambridge University Press}
}

@article{karyappa2014breakup,
  title={Breakup of a conducting drop in a uniform electric field},
  author={Karyappa, Rahul B and Deshmukh, Shivraj D and Thaokar, Rochish M},
  journal={Journal of fluid mechanics},
  volume={754},
  pages={550--589},
  year={2014},
  publisher={Cambridge University Press}
}

@article{wagoner2021electrohydrodynamics,
  title={Electrohydrodynamics of lenticular drops and equatorial streaming},
  author={Wagoner, Brayden W and Vlahovska, Petia M and Harris, Michael T and Basaran, Osman A},
  journal={Journal of Fluid Mechanics},
  volume={925},
  pages={A36},
  year={2021},
  publisher={Cambridge University Press}
}

@article{wang2024lattice,
  title={Lattice Boltzmann modelling and study of droplet equatorial streaming in an electric field},
  author={Wang, Geng and Lei, Timan and Yang, Junyu and Fei, Linlin and Chen, Jin and Luo, Kai H},
  journal={Journal of Fluid Mechanics},
  volume={988},
  pages={A40},
  year={2024},
  publisher={Cambridge University Press}
}

@article{zhang20052d,
  title={A 2D lattice Boltzmann study on electrohydrodynamic drop deformation with the leaky dielectric theory},
  author={Zhang, Junfeng and Kwok, Daniel Y},
  journal={Journal of Computational Physics},
  volume={206},
  number={1},
  pages={150--161},
  year={2005},
  publisher={Elsevier}
}

@article{tomar2007two,
  title={Two-phase electrohydrodynamic simulations using a volume-of-fluid approach},
  author={Tomar, Gaurav and Gerlach, Daniel and Biswas, Gautam and Alleborn, Norbert and Sharma, Ashutosh and Durst, Franz and Welch, Samuel WJ and Delgado, Antonio},
  journal={Journal of Computational Physics},
  volume={227},
  number={2},
  pages={1267--1285},
  year={2007},
  publisher={Elsevier}
}

@article{hua2008numerical,
  title={Numerical simulation of deformation/motion of a drop suspended in viscous liquids under influence of steady electric fields},
  author={Hua, Jinsong and Lim, Liang Kuang and Wang, Chi-Hwa},
  journal={Physics of Fluids},
  volume={20},
  number={11},
  year={2008},
  publisher={AIP Publishing}
}

@article{paknemat2012numerical,
  title={Numerical simulation of drop deformations and breakup modes caused by direct current electric fields},
  author={Paknemat, H and Pishevar, AR and Pournaderi, P},
  journal={Physics of Fluids},
  volume={24},
  number={10},
  year={2012},
  publisher={AIP Publishing}
}

@article{nganguia2015immersed,
  title={An immersed interface method for axisymmetric electrohydrodynamic simulations in Stokes flow},
  author={Nganguia, H and Young, Y-N and Layton, AT and Hu, W-F and Lai, M-C},
  journal={Communications in Computational Physics},
  volume={18},
  number={2},
  pages={429--449},
  year={2015},
  publisher={Cambridge University Press}
}

@article{das2017electrohydrodynamics,
  title={Electrohydrodynamics of viscous drops in strong electric fields: numerical simulations},
  author={Das, Debasish and Saintillan, David},
  journal={Journal of Fluid Mechanics},
  volume={829},
  pages={127--152},
  year={2017},
  publisher={Cambridge University Press}
}

@article{liu2019phase,
  title={A phase-field-based lattice Boltzmann modeling of two-phase electro-hydrodynamic flows},
  author={Liu, Xi and Chai, Zhenhua and Shi, Baochang},
  journal={Physics of Fluids},
  volume={31},
  number={9},
  year={2019},
  publisher={AIP Publishing}
}

@article{ramaswamy1999deformation,
  title={The deformation of a viscoelastic drop subjected to steady uniaxial extensional flow of a Newtonian fluid},
  author={Ramaswamy, S and Leal, LG},
  journal={Journal of non-newtonian fluid mechanics},
  volume={85},
  number={2-3},
  pages={127--163},
  year={1999},
  publisher={Elsevier}
}

@article{hooper2001transient,
  title={Transient polymeric drop extension and retraction in uniaxial extensional flows},
  author={Hooper, Russell W and De Almeida, Valmor F and Macosko, Christopher W and Derby, Jeffrey J},
  journal={Journal of non-newtonian fluid mechanics},
  volume={98},
  number={2-3},
  pages={141--168},
  year={2001},
  publisher={Elsevier}
}

@article{aggarwal2007deformation,
  title={Deformation and breakup of a viscoelastic drop in a Newtonian matrix under steady shear},
  author={Aggarwal, Nishith and Sarkar, Kausik},
  journal={Journal of Fluid Mechanics},
  volume={584},
  pages={1--21},
  year={2007},
  publisher={Cambridge University Press}
}

@article{ha1999deformation,
  title={Deformation and breakup of a second-order fluid droplet in an electric field},
  author={Ha, Jong-Wook and Yang, Seung-Man},
  journal={Korean Journal of Chemical Engineering},
  volume={16},
  pages={585--594},
  year={1999},
  publisher={Springer}
}

@article{ha2000deformation,
  title={Deformation and breakup of Newtonian and non-Newtonian conducting drops in an electric field},
  author={Ha, Jong-Wook and Yang, Seung-Man},
  journal={Journal of Fluid Mechanics},
  volume={405},
  pages={131--156},
  year={2000},
  publisher={Cambridge University Press}
}

@article{lima2014numerical,
  title={Numerical simulation of electrohydrodynamic flows of Newtonian and viscoelastic droplets},
  author={Lima, NC and d’Avila, MA},
  journal={Journal of Non-Newtonian Fluid Mechanics},
  volume={213},
  pages={1--14},
  year={2014},
  publisher={Elsevier}
}

@article{zhao2025electrohydrodynamic,
  title={Electrohydrodynamic effects on the viscoelastic droplet deformation in shear flows},
  author={Zhao, Jiachen and Dzanic, Vedad and Wang, Zhongzheng and Sauret, Emilie},
  journal={Physics of Fluids},
  volume={37},
  number={1},
  year={2025},
  publisher={AIP Publishing}
}

@article{alves2021numerical,
  title={Numerical methods for viscoelastic fluid flows},
  author={Alves, MA and Oliveira, PJ and Pinho, FT},
  journal={Annual Review of Fluid Mechanics},
  volume={53},
  number={1},
  pages={509--541},
  year={2021},
  publisher={Annual Reviews}
}

@article{fattal2004constitutive,
  title={Constitutive laws for the matrix-logarithm of the conformation tensor},
  author={Fattal, Raanan and Kupferman, Raz},
  journal={Journal of Non-Newtonian Fluid Mechanics},
  volume={123},
  number={2-3},
  pages={281--285},
  year={2004},
  publisher={Elsevier}
}

@article{lopez2019adaptive,
  title={An adaptive solver for viscoelastic incompressible two-phase problems applied to the study of the splashing of weakly viscoelastic droplets},
  author={L{\'o}pez-Herrera, Jos{\'e}-Mar{\'\i}a and Popinet, St{\'e}phane and Castrej{\'o}n-Pita, Alfonso-Arturo},
  journal={Journal of Non-Newtonian Fluid Mechanics},
  volume={264},
  pages={144--158},
  year={2019},
  publisher={Elsevier}
}

@article{lopez2011charge,
  title={A charge-conservative approach for simulating electrohydrodynamic two-phase flows using volume-of-fluid},
  author={L{\'o}pez-Herrera, JM and Popinet, St{\'e}phane and Herrada, MA2764018},
  journal={Journal of Computational Physics},
  volume={230},
  number={5},
  pages={1939--1955},
  year={2011},
  publisher={Elsevier}
}

@article{DAS2026105633,
title = {Effect of viscoelasticity on electrohydrodynamic drop deformation},
journal = {International Journal of Multiphase Flow},
volume = {197},
pages = {105633},
year = {2026},
issn = {0301-9322},
author = {Santanu Kumar Das and Sarika Shivaji Bangar and Amaresh Dalal and Gaurav Tomar}
}

@article{popinet2015quadtree,
  title={A quadtree-adaptive multigrid solver for the Serre--Green--Naghdi equations},
  author={Popinet, St{\'e}phane},
  journal={Journal of Computational Physics},
  volume={302},
  pages={336--358},
  year={2015},
  publisher={Elsevier}
}

\end{document}